\documentclass{LMCS}

\usepackage{amsmath,amstext,amssymb,enumerate,xspace}
\usepackage{pst-tree,url,amsthm}
\usepackage{hyperref}

\psset{levelsep=0.7cm,treemode=D, nodesepB=2pt,nodesepA=2pt}

\def\doi{2 (3:1) 2006}
\lmcsheading%
{\doi}
{30}
{}
{}
{Aug.~18, 2005}
{Jul.~26, 2006}
{}

\begin{document}

\title[Complexity of XPath containment]{On the complexity of XPath
  containment in the presence of disjunction, DTDs, and variables}

\author[F.\ Neven]{Frank Neven\rsuper a}
\address{{\lsuper a}Hasselt University and
Transnational University of Limburg}
\email{frank.neven@uhasselt.be}

\author[T.\ Schwentick]{Thomas Schwentick\rsuper b}
\address{{\lsuper b}University of Dortmund}
\email{thomas.schwentick@udo.edu}


\newtheorem{theorem}{Theorem}[section]
\newtheorem{definition}[theorem]{Definition}
\newtheorem{lemma}[theorem]{Lemma}
\newtheorem{corollary}[theorem]{Corollary}
\newtheorem{proposition}[theorem]{Proposition}
\newtheorem{example}[theorem]{Example}
\newtheorem{claim}[theorem]{Claim}

\newcommand{\NC}{\newcommand}

\newlength{\algwidth}
\setlength{\algwidth}{\linewidth}
\addtolength{\algwidth}{-2cm}

\renewcommand{\algo}[1]
{\begin{center}\fbox{\begin{minipage}{\algwidth}#1\end{minipage}}\end{center}}

\NC{\mytrue}{\mbox{\tt true}\xspace}
\NC{\myfalse}{\mbox{\tt false}\xspace}

\NC{\vars}{\text{vars}\xspace}
\NC{\evars}{\text{evars}\xspace}
\NC{\xvars}{\text{xvars}\xspace}
\NC{\disjunction}{\text{$|$}\xspace}
\NC{\ineq}{\text{$\neq$}\xspace}
\NC{\dtd}{\text{DTD}\xspace}
\NC{\wildcard}{\text{$*$}\xspace}
\NC{\descendant}{\text{$/\!/$}\xspace}
\NC{\filter}{\text{\rm[\,]}\xspace}
\NC{\filterAnd}{\text{and}\xspace}
\NC{\filterOr}{\text{or}\xspace}
\NC{\child}{\text{$/$}\xspace}

\NC{\tself}{{\tt self}\xspace}
\NC{\tchild}{{\tt child}\xspace}
\NC{\tdescendant}{{\tt descendant}\xspace}
\NC{\tdescendantself}{{\tt descendant-or-self}\xspace}
\NC{\xcont}{{\sc XContainment}\xspace}
\NC{\xdcont}{{\sc XDContainment}\xspace}
\NC{\xfcont}{{\sc XFContainment}\xspace}

\NC{\sub}{\text{Sub}}

\NC{\I}{\text{\tt I}\xspace}
\NC{\II}{\text{\tt II}\xspace}

\NC{\XP}{\ensuremath{\text{\rm XP}}\xspace}
\NC{\XPfull}{XP(\child,\descendant,\filter,\wildcard,\disjunction)\xspace}

\NC{\defem}[1]{{\bf #1}}

\maketitle

\NC{\disj}{\text{disj}}

\NC{\dom}{\text{dom}}%
\NC{\Dom}{\text{dom}}%

\NC{\sem}[1]{[\![ #1 ]\!] }%
\NC{\semt}[1]{[\![ #1 ]\!]_t }%
\NC{\val}{\text{val}}%
\NC{\lab}{\text{lab}}%
\NC{\data}{{\bf D}}
\NC{\bt}{t}
\NC{\bft}{t}
\NC{\subs}{\subseteq}

\NC{\none}{{\tt none}\xspace}
\renewcommand{\epsilon}{\varepsilon}


\begin{abstract}
  XPath is a simple language for navigating an XML-tree and returning
  a set of answer nodes.  The focus in this paper is on the complexity
  of the containment problem for various fragments of XPath.  We
  restrict attention to the most common XPath expressions which
  navigate along the child and/or descendant axis.  In addition to
  basic expressions using only node tests and simple predicates, we
  also consider disjunction and variables (ranging over
  nodes). Further, we investigate the containment problem relative to
  a given DTD.  With respect to variables we study two semantics, (1)
  the original semantics of XPath, where the values of variables are
  given by an outer context, and (2) an existential semantics
  introduced by Deutsch and Tannen, in which the values of variables
  are existentially quantified. In this framework, we establish an
  exact classification of the complexity of the containment problem
  for many XPath fragments.
\end{abstract}

\section{Introduction}
XPath is a simple language for navigating an XML document and
selecting a set of element nodes~\cite{xpath}. At the same time it is
also the main XML selection language. Indeed, XPath expressions are
used, for instance, as basic patterns in several XML query languages
like XQuery \cite{xquery} and XSLT \cite{bmnjournal,xslt}; they are
used in XML Schema to define keys~\cite{w3schema}, and in XLink
\cite{xlink} and XPointer \cite{xpointer} to reference elements in
external documents.  In every such context an instance of the
containment problem is present: optimizing XPath expressions can be
accomplished by an algorithm for containment, and XSLT rule selection
and inference of keys based on XPath expressions again reduces to
containment.  In this article we focus on the complexity of the
containment problem of various fragments of XPath 1.0 using only the
most common axes, \child and \descendant, and extensions in which
variables can refer to data values. Furthermore, the containment
problem relative to a given DTD is investigated. In all cases, we only
consider the Boolean containment problem. Here, given two XPath
expressions $p$ and $q$, the problem asks whether the fact that $p$
selects some path from the root to a vertex implies that also $q$
selects some path from the root to a vertex. This restriction is
justified as all complexity results we obtain easily transfer to unary
and binary containment in the spirit of Proposition 1 in
\cite{miklaujacm}.

The XPath containment problem already attracted quite some
attention~\cite{dtxpath,miklaujacm,moerkotte,wooddood,woodicdt,marxedbt}.
We next discuss the known results together with our own
contributions.

A general result establishing a strong upper bound for a large
fragment of XPath is due to Marx presented in \cite{marxedbt}. It is
shown there that the containment problem for navigational XPath,
allowing navigation along all axes, even relative to a DTD, is in {\sc
  exptime}.

Other work has concentrated, like this article, on XPath expressions
that can only navigate downwards in an XML tree and do not use the
order between siblings, i.e., the fragment using only the \child and
\descendant axis. Different fragments can be defined by
allowing or disallowing the use of the wild-card $*$ in node tests,
and filter predicates in location steps. In the spirit of the abbreviated
syntax of XPath we use \child to indicate the use of the \tchild axis,
\descendant for the \tdescendant axis, \wildcard for the wild-card and
\filter for filter predicates. We denote XPath fragments by listing
the allowed operators. For 
instance, XP(\child,\descendant,\filter) denotes the XPath fragment
with the \tchild and \tdescendant axes in which the use of filter
predicates is allowed, but no wild-cards in node tests. 

Among other results, Miklau and Suciu \cite{miklaujacm} obtain that
containment for XP(\child,\descendant, \filter,\wildcard) is {\sc
  conp}-complete~\cite{miklaujacm}. Here, inside filter predicates and
between location steps, no Boolean operators are allowed. 

\paragraph{Contributions.}

The first family of fragments we consider is obtained by allowing
disjunction (\disjunction) in filter predicates and in location steps. 
We show that, in principle, adding disjunction to
XP(\child,\descendant, \filter,\wildcard) does not make the
containment problem harder. Surprisingly, when the set of allowed
element names (labels) in XML documents is restricted, and given as
part of the input then the containment problem becomes much harder:
complete for {\sc pspace}. The results on fragments with disjunction
are shown in Table \ref{tab:disjunction}

\begin{table}[ht]
\begin{center}  
\begin{tabular}{|c|c|c|c|c|l|c|}
\hline \child & \descendant & \filter & \disjunction & \wildcard & 
Complexity & Reference \\
\hline\hline
+& +& +& & +&\textsc{conp}-complete & \cite{miklaujacm}\\
+& +& +& +& +&\textsc{conp}-complete
&(\ref{theo:disjunctionlower},\ref{theo:disjunctionupper})\\ 
+& & & +& &\textsc{conp}-complete&(\ref{theo:disjunctionlower},\ref{theo:disjunctionupper})\\
& +& & +& &\textsc{conp}-complete&(\ref{theo:disjunctionlower},\ref{theo:disjunctionupper})\\\hline
+& +& +& +&+ &\textsc{pspace}-complete (given alphabet)&(\ref{theo:finitelower},\ref{theo:finiteupper})\\
+& +& & +&&\textsc{pspace}-complete (given alphabet)&(\ref{theo:finitelower},\ref{theo:finiteupper})\\
\hline
 \end{tabular}
\end{center}
\caption{\label{tab:disjunction} The complexity of containment for expressions
  with disjunction. Square brackets refer to the references,
  parentheses to results of this article.} 
\end{table}

Deutsch and Tannen \cite{dtxpath} consider XPath containment in the
presence of DTDs and Simple XPath Integrity Constraints (SXICs)
\cite{dtxpath}. Here, the input to the containment problem consists of
two XPath expressions $p$, $q$, a DTD and/or a set of integrity
constraints and it is asked whether $p$ selects a subset of the
elements that $q$ selects, in all documents respecting the DTD and/or
the constraints.  They show that this problem is undecidable in
general and in the presence of bounded SXICs and DTDs. When only DTDs
are present they have a {\sc pspace} lower bound and leave the exact
complexity as an open question.

We indicate the presence of a DTD by XP(DTD,...).  We give a simple
proof that containment testing for
XP(DTD,\child,\descendant,\filter,\wildcard,\disjunction) is in {\sc
  exptime} (although this result is covered by the above mentioned
result of Marx \cite{marxedbt}) and obtain that containment for
XP(DTD,\child,\descendant,\disjunction) and for
XP(DTD,\child,\descendant,\filter,\wildcard) are hard for {\sc
  exptime}. We also study the complexity of more restrictive fragments
in the presence of DTDs. It turns out that containment of
XP(DTD,\child,\descendant) is in {\sc PTime}.  On the other hand,
containment of XP(DTD,\child,\filter) is {\sc conp}-complete and
containment of XP(DTD,\descendant,\filter) is {\sc conp}-hard. It is
not clear whether or how the upper bound proof in the former case can
be extended to include, for instance, the descendant operator. The
results about the containment problem in the presence of DTDs are
summarized in Table \ref{tab:dtd}.

\begin{table}[ht]
\begin{center}  
\begin{tabular}{|c|c|c|c|c|c|l|c|}
\hline \dtd&\child & \descendant & \filter & \disjunction & \wildcard & 
Complexity & Reference  \\
\hline\hline
+&+&+  &  & &  &in \textsc{p}&(\ref{theo:dtd:p})\\\hline
+& +& & +& & &\textsc{conp}-complete&(\ref{theo:dtd:npupper},\ref{theo:dtd:nplower},\cite{woodwebdb})\\
+& &+ & +& & &\textsc{conp}-hard&(\ref{theo:dtd:nplower},\cite{woodwebdb})\\\hline
+& + & + &+&  + & +& \textsc{exptime}-complete&(\ref{theo:dtd:exptimeupper},\ref{theo:dtd:exptimelower})\\
+& + & + &  & + &  & \textsc{exptime}-complete &(\ref{theo:dtd:exptimeupper},\ref{theo:dtd:exptimelower})\\
\hline
 \end{tabular}
\end{center}
\caption{\label{tab:dtd} The complexity of containment in the presence 
of DTDs.}
\end{table}

The XPath recommendation allows variables to be used in XPath
expressions on which equality tests can be performed. For instance,
$//a[\$x=@b][\$y\neq @c]$ selects all $a$-descendants whose
$b$-attribute equals the value of variable $\$x$ and whose
$c$-attribute differs from the value of variable $\$y$. However, under
the XPath semantics the value of all variables should be specified by
the outer context (e.g., in the XSLT template in which the pattern is
issued). We indicate the use of variables with XPath semantics by
XP(...,\xvars,...). 
So the semantics of an XPath expression is defined with respect to a variable
mapping. We show that the complexity of containment is {\sc
  pspace}-complete under this semantics. For the lower bound, it
suffices to observe that with variables a finite alphabet can be
simulated. We obtain the upper bound by reducing the containment
problem to the containment of several expressions without variables.

In addition to the XPath semantics, Deutsch and Tannen \cite{dtxpath}
considered an existential semantics for variables: an expression matches a
document if there {\em exists} a suitable assignment for the
variables.  We denote variables with existential semantics by 
XP(...,\evars,...). 
In \cite{dtxpath} it is shown that containment of
XP(\child,\descendant,\filter,\wildcard,\evars) and
XP(\child,\descendant,\filter,\disjunction,\evars) is $\Pi_2^P$-hard,
and that containment of XP(\child,\descendant,
\filter,\disjunction,\evars) under fixed bounded SXICs is in $\Pi_2^P$.
We extend their result by showing that containment of
XP(\child,\descendant,\filter,\disjunction,\evars,\ineq), that is,
inequality tests on variables and attribute values are allowed,
remains in $\Pi_2^P$. 
Surprisingly, the further addition of \wildcard to this fragment makes
the containment problem undecidable. The results about XPath
containment for fragments with variables are indicated in Table
\ref{tab:variables}

\begin{table}[htbp]
\begin{center}  
\begin{tabular}{|c|c|c|c|c|c|c|c|l|c|}
\hline \child & \descendant & \filter  & \wildcard & \disjunction &
\xvars & \evars & \ineq & complexity & Reference  \\
\hline\hline
+&&+  &  & &+ && + & \textsc{conp}-hard&(\ref{theo:xvars:lower})\\
+&+&  &  & &+ && + & \textsc{conp}-hard&(\ref{theo:xvars:lower})\\
+&+&+  & + &+ &+ && + & \textsc{pspace}-complete&(\ref{theo:xvars:upper},\ref{theo:xvars:lower})\\
+&+&  & + &+ &+ && + & \textsc{pspace}-complete&(\ref{theo:xvars:upper},\ref{theo:xvars:lower})\\
\hline
+&+&+  &  & &&+&  & \textsc{conp}-complete&\cite{dtxpath}\\
+&+&   &  & &&+&  & \textsc{conp}-complete&\cite{dtxpath},
(\ref{theo:evars:lower})\\\hline 
+&+&+  &+ &+&&+&  & \textsc{$\Pi_2^p$}-complete & \cite{dtxpath}\\
+& &  &  &+ &&+& & \textsc{$\Pi_2^p$}-complete & \cite{dtxpath}, (\ref{theo:evars:lower})\\
+& &+  &  & &&+&+ & \textsc{$\Pi_2^p$}-complete &
(\ref{theo:evars:lower},\ref{theo:evars:upper})\\ 
+&+&+  &  &+&&+&+  & \textsc{$\Pi_2^p$}-complete&
(\ref{theo:evars:lower},\ref{theo:evars:upper})\\\hline
+&+&+  &+  &+ &&+&+  & undecidable& (\ref{theo:evars:undec})\\\hline
 \end{tabular}
\end{center}
\caption{\label{tab:variables} The complexity of containment in the presence 
of variables. Note that \xvars and \evars refer to the original
XPath semantics and to existential semantics, respectively.}
\end{table}

\paragraph{Further related work.}

In \cite{woodicdt}, Wood shows that containment of
XP(\child,\descendant,\filter,\wildcard) in the presence of DTDs is
decidable. He also studies conditions for which containment under DTDs
is in {\sc ptime}.  Benedikt, Fan, and Kuper study the expressive
power and closure properties of fragments of
XPath~\cite{benedictfankuper}. They also consider sound and complete
axiom systems and normal forms for some of these fragments.  Hidders,
and Benedikt, Fan, and Geerts considered the complexity of
satisfiability of XPath
expressions~\cite{hiddersxpath,florisxpath,DBLP:conf/dbpl/GeertsF05}.
The complexity of XPath evaluation has been studied by Gottlob, Koch,
and Pichler in \cite{kochlics,kochvldb}, while its expressive power
has been addressed by Marx in \cite{MarxRSemantic05} and
\cite{marxicdt}. This article is based on \cite{NevenSXPath03}.

\paragraph{Organization.}

This article is organized as follows. In Section~\ref{sec:defs}, we
define DTDs and the basic XPath fragments. We also introduce the
necessary machinery w.r.t. unranked tree automata.  In
Section~\ref{sec:disjunction}, \ref{sec:dtd}, and \ref{sec:variables}
we consider disjunction, DTDs, and variables, respectively. We
conclude in Section~\ref{sec:discussion}.


\section{Preliminaries}
\label{sec:defs}

In this section, we define the tree abstraction of XML documents, DTDs
and the fragments of XPath that we consider.

\subsection{XML-trees}
For the rest of this paper, we fix an infinite set $\Sigma$ of labels
and an infinite set $\data$ of data values. Only in
Section~\ref{sec:finite}, we consider XPath expressions over a finite
alphabet.  The set $A$ is always a finite set of attributes.  An XML
document is faithfully modeled by a finite unranked tree with labels
from $\Sigma$ in which the attributes of the nodes have $\data$-values
and in which the children of each node are ordered.

It is common to model the underlying tree as a {\em tree domain}. To
this end, the edges connecting a node with its children are numbered
from 1 to $n$, according to the ordering of its children. Each path
from the root to a node then corresponds to a sequence of numbers.
Finally, each node is identified with this sequence. In particular,
the root, which corresponds to the document node \cite{xpath}, is
represented by the empty string denoted by $\epsilon$. 

More formally, a \defem{tree domain} $D$ is a finite subset of
${\Bbb{N}}^*$ with the following closure properties:
\begin{itemize}
\item If $v\cdot i\in D$, where $v\in
{\Bbb{N}}^*$ and $i\in {\Bbb{N}}$, then $v\in D$.
\item  If $i>1$ and $v\cdot i\in D$, then also $v \cdot(i-1)\in D$.
\end{itemize}
We call the elements of $D$ \defem{vertices}. A vertex $vi$ with
$i\in{\Bbb{N}} $ is a \defem{child} of a vertex $v$.  Conversely, $v$
is called the \defem{parent} of $vi$. A vertex $vu$ with $u \in
{\Bbb{N}^+}$ is a \defem{descendant} of $v$. We also say that $v$ is
an \defem{ancestor} of $vu$.

\begin{definition}\rm
  An \defem{XML-tree\/} (tree for short) is a triple
  $\bt=(\dom(\bft),\lab_\bft,\lambda_t)$, where $\dom(\bft)$ is a tree
  domain over $\Bbb{N}$, and $\lab_\bft:\dom(\bft)\to \Sigma$ and,
  for each $a\in A$, $\lambda_t^ a:\dom(t)\to \data$ are partial functions.
  Intuitively, $\lab_t(v)$ is the label of $v$, while $\lambda_t^a(v)$
  is the value of $v$'s $a$-attribute, if it has one.
\end{definition}

Of course, in real XML documents there can be vertices with mixed
content, but these can easily be modeled by using auxiliary
intermediate nodes as explained in \cite{bmnjournal}.  We consider
attributes only in Section~\ref{sec:variables}.

For a vertex $v\in\dom(t)$, we denote by $t_v$ the \defem{sub-tree} of
$\dom(t)$ rooted at $v$. As a tree domain in itself, this is the set
$\{w\mid vw\in\dom(t)\}$.

\subsection{DTDs}
We formalize Document Type Definitions (DTDs) as context-free grammars
with regular expressions on the right-hand side of rules. As usual, we
denote by $L(r)$ the language defined by the regular expression $r$.

\begin{definition}\rm 
A \defem{DTD} is a tuple $(d,S_d,\Sigma_d)$ where 
$\Sigma_d$ is a finite subset of $\Sigma$,
$S_d\in \Sigma_d$ is the
start symbol, and $d$ is a mapping from $\Sigma_d$ to
the set of regular expressions over $\Sigma_d$. 
A tree $t$ \defem{matches} a DTD $d$ iff $\lab_t(\varepsilon)
=S_d$ and for every $u\in\dom(t)$ with $n$ children, $\lab_t(u1)
\cdots\lab_t(un)\in L(d(\lab(u)))$. We denote by $L(d)$ the
set of all trees that match $d$.
\end{definition}

Note that DTDs do not constrain the value of attributes in any way.
We usually refer to a DTD by $d$ rather than by $(d,S_d,\Sigma_d)$.

\subsection{XPath}
\label{sec:xpath}
We next define the core fragment of XPath that we will consider in
Sections~\ref{sec:disjunction} and \ref{sec:dtd}.  In our definition
we follow Marx~\cite{marxpods}. In Section \ref{sec:variables}, we
consider a larger fragment which allows the use of attribute values.

\begin{definition}\rm \label{def:syn}
An \defem{XPath expression} is generated by the following grammar:
\begin{center}
\begin{tabular}{rcl}
lpath & $::=$&  lstep $\mid$ lpath '/'
lpath $\mid$ lpath '$|$' lpath\\
lstep & $::=$ & axis '::' node-test ('['fexpr']')$^*$\\
axis & $::=$ & \tself $\mid$ \tchild $\mid$ \tdescendant\\
fexpr & $::=$ & lpath $\mid$ lpath 'or' lpath\\
\end{tabular}
\end{center}
Here, lpath is the start symbol which is short for location path;
node-test is either a label or the wild-card '$*$'. 
\end{definition}

We write $|p|$ for the \defem{size} of an XPath expression, which is the total
number of occurrences in $p$ of axes \tchild and \tdescendant
(including those in filter expressions). 

We use $\bigcup$ to denote a big disjunction of expressions.

Note that, as we only consider expressions which navigate downwards in
the tree, we do not allow {\em absolute location paths}, i.e., paths
requiring to be evaluated from the root.  For convenience, we further
assume that location steps with the \tself-axis only occur at the
top-level. This is no loss of generality, as one can always translate
a location path of the form
$$\text{axis}::\sigma[e_1]\cdots[e_k]
/\tself::\sigma'[f_1]\cdots[f_\ell]$$ into
$$\text{axis}::\gamma[e_1]\cdots[e_k] [f_1]\cdots[f_\ell]$$ where
$$\gamma = \left\{ \begin{array}{ll} 
\sigma & \text{if $\sigma'=*$; and,}\\
\sigma' & \text{if $\sigma=\sigma'$ or $\sigma=*$}.
\end{array} \right.
$$
Of course, when $\sigma\not=\sigma'$ are two labels in the above
expression, then it is unsatisfiable.

For notational brevity, we often use abbreviated syntax for XPath
expressions \cite{xpath}. Thus, instead of
$$\text{{\tchild :: $a$ [\tchild :: $b$][\tdescendant :: e]/\tdescendant :: $c$}},$$
we simply write $a [b][.\descendant e]\descendant c$. Note that the
sub-expression $.\descendant e$ is the abbreviated notation for
$\text{\tdescendantself :: $*$ / \tchild :: e}$ and
  thus accounts for \tdescendant :: e.

For the definition of the semantics of XPath expressions we again
basically follow \cite{marxpods}.  

\begin{definition}\rm \label{def:sem}
  For each tree $t$, each location path $p$ induces a binary relation
  $\semt{p}$ which is inductively defined as follows:
\begin{itemize}
\item $\semt{a :: n [e_1]\cdots[e_k]}$ is the set of all pairs $(u,v)$
  for which all the following conditions hold:
  \begin{itemize}
  \item if $a$ is \tchild then $v$ is a child of $u$;
  \item if $a$ is \tdescendant then $v$ is a descendant of $u$;
  \item if $a$ is \tself then $v=u$;
  \item if $n$ is a label then $\lab_t(v)=n$, otherwise $n$ is $*$ and
  the label of $v$ can be arbitrary; and, 
\item $E_t(v,e_i)$ is true, for each $i\le k$.  Here, $E_t$ is defined
  as follows:
\begin{itemize}
\item $E_t(v,p)$ is true for a vertex $v$ and a location path $p$ if
  and only if there is a vertex $w$ such that $(v,w)\in\semt{p}$.
\item $E_t(v,e_1 \text{ or } e_2)$ is true for a vertex $v$ if and only
  if $E_t(v,e_1)$ or $E_t(v,e_2)$ is true.
\end{itemize}

  \end{itemize}
\item $\semt{p/q} = \semt{p} \circ \semt{q}$ (where $\circ$ denotes the
  composition of binary relations); and,
\item $\semt{p \mid q}=\semt{p} \cup\semt{q}$.
\end{itemize}

\end{definition}

So, the semantics definition associates with every tree $t$ and every
expression $p$ a binary relation. When the {\em context vertex}, i.e.,
the first vertex in pairs, is fixed to be the root then every
expression defines the set $\{v \mid (\epsilon,v)\in \semt{p}\}$.
Recall that $\epsilon$ denotes the root of a tree. We say  a tree $t$
\defem{matches} an expression $p$ (written: $t\models p$) if there is
some vertex $v$ in $t$ such that 
$(\epsilon,v)\in\semt{p}$. In the latter case we interpret an
expression $p$ as a \defem{Boolean query}.

\begin{table}[htbp]
  \centering
  \begin{tabular}{|c|c|}
\hline
Symbol & Meaning\\
\hline
\child & child axis is allowed\\
\descendant & descendant axis is allowed\\
\filter & filter expressions are allowed\\
\disjunction & disjunction ('or' and $|$) is allowed\\
\wildcard & wild-cards are allowed\\
\hline
  \end{tabular}
  \caption{Symbols used in the notation for XPath fragments. }
  \label{tab:not}
\end{table}

We denote sub-fragments of the above defined XPath fragment using the
abbreviated syntax. We use the notations explained in Table
\ref{tab:not}.  If filter expressions are not allowed, location steps
are only of the form axis :: node-test. Disjunction is allowed in
location paths and in filter expressions. If wild-cards are not
allowed, every node-test has to be a label.

We denote XPath fragments by XP(...) where inside the brackets the
allowed features are listed. For instance, we write
XP(\child,\filter,\disjunction) for the fragment, where the wild-card is
not allowed and the descendant axis can not be used.

\begin{figure}[htbp]
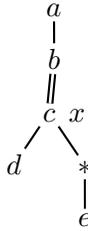

  \centering
  \pstree{\TR{$a$}}
  {\pstree{\TR{$b$}}
   {\psset{doubleline=true}
    {\pstree{\TR{$c$}~[tnpos=r]{$x$}}
     {\psset{doubleline=false}
      \TR{$d$}
      \pstree{\TR{$*$}}
      {\TR{$e$}}
     }
    }
   } 
  }
  \caption{The tree pattern corresponding to $a\child b\descendant c[d][*/e]$}
  \label{fig:treepat}
\end{figure}

In some proofs, we view expressions $p$ from
XP(\child,\descendant,\filter,\wildcard) as \defem{tree patterns} as
described by Miklau and Suciu~\cite{miklaujacm}.  For example, the
expression $a\child b\descendant c[d][*/e]$ corresponds to the tree
pattern in Figure~\ref{fig:treepat}.  Single edge and double edge
correspond to the child and descendant axis, respectively.  
We denote the tree pattern associated with an expression $p$ by
$\tau(p)$. 

From this point of view, $t\models p$ if and only if
there is a homomorphism $h$ from $\tau(p)$ to $t$, i.e., $h$ maps the
nodes of $h(p)$ to the nodes of $t$ such that (1) $h(v)$ has the same
label as $v$ unless $v$ carries a wild-card, (2) $h(v)$ is a child
(descendant) of $h(u)$ if and only if $v$ is a child (descendant) of
$u$.  So, $h$ respects labels, child and descendant, and does not care
about $*$. 

Every tree pattern has one selecting node: all
pairs $(u,v)$ of nodes of the input tree are selected, for which the
root of the tree pattern can be mapped to $u$ and the selecting node to $v$.
In Figure~\ref{fig:treepat}, the selecting node is labeled 
by $x$.

We will frequently make use of the fact that every expression $p$ from
\XPfull can be written in \defem{disjunctive normal form}, i.e., in
the form $p_1 \mid \cdots \mid p_n$, where each $p_i$ is an expression
from XP(\child,\descendant,\filter,\wildcard). It should be noted that
$n$ can be exponential in $|p|$.

\subsection{Containment}
\label{sec:def:cont}

Corresponding to the three different interpretations of an XPath
expression as a binary, unary or Boolean query, there are three
different notions of query containment.

\begin{definition}\rm
For two XPath expressions $p$ and $q$,
\begin{itemize}

\item  $p$ is \defem{contained as a binary query} in $q$, denoted
by $p \subs_2 q$, if $\semt{p}\subs\semt{q}$, for every tree $t$;

\item $p$ is \defem{contained as a unary query} in $q$, denoted by
$p \subs_1 q$ if, for each tree $t$ and each vertex $v$ of $t$,
$(\epsilon,v)\in\semt{p}$ implies $(\epsilon,v)\in\semt{q}$; and

\item $p$ is \defem{contained as a Boolean query} in $q$, denoted
  $p\subs q$, if, for every tree $t$, $t\models p$ implies $t\models q$.
\end{itemize}
\end{definition}

\begin{definition}\rm
  \xcont is the algorithmic problem to decide for two XPath
  expressions $p$ and $q$, whether $p\subs q$.
\end{definition}

For a DTD $d$ and two XPath expressions $p$ and $q$, by $p\subs_d q$,
we denote that $t\models p$ implies $t\models q$, for all trees $t$
that match $d$.

\begin{definition}\rm
  \xdcont is the algorithmic problem to decide for two XPath
  expressions $p$ and $q$ and a DTD, whether $p\subs_d q$.
\end{definition}

The restriction to Boolean containment is justified:
it is shown in \cite{miklaujacm} that for
XP(\child,\descendant,\filter,\wildcard) the complexity of deciding
binary (or $k$-ary, for any $k$) containment is the same as the
complexity of deciding Boolean containment. For binary queries, this result can be
generalized to all fragments we consider in this article. More
precisely: all complexity results stated in this article hold also for
unary and binary containment. First of
all, binary and unary containment are computationally equivalent in our framework as
from a node $v$ expressions can only navigate inside the subtree
induced by $v$. Furthermore, to get from unary to Boolean containment,
it is easy to see that the result in  \cite{miklaujacm} only requires
the child axes, i.e., only one node is added as a child of the
selecting node of the tree pattern. It is straightforward to get the
same statement also in the presence of a DTD. For the fragment
XP(\descendant,\disjunction) a similar approach works: for the label $a$
of the selecting node a new label $a'$ is introduced and every
reference of a not-selecting node to $a$ is replaced by $a | a'$. 
The lower bound we prove for XP(\descendant,\filter) also goes through
for the binary and unary case.

\subsection{Unranked tree automata}

We recall the definition of non-deterministic tree automata over
unranked trees from \cite{bmw}. These are used in the proofs of
Theorem~\ref{theo:dtd:p} and Theorem~\ref{theo:dtd:exptimeupper}.  We
refer the unfamiliar reader to \cite{nevenSR} for a gentle
introduction.  The alternating automata of Section~\ref{sec:finite}
operate over ranked trees.

\begin{definition}\rm
  A \defem{nondeterministic tree automaton (NTA)\/} is a tuple
  $A=(Q,\Delta, \allowbreak \delta,F)$, where $Q$ is a finite set of
  states, $\Delta$ is a finite alphabet, $F\subseteq Q$ is the set of
  final states, and $\delta$ is a function $\delta: Q\times\Delta\to
  2^{(Q^*)}$ such that $\delta(q,a)$ is a regular string language over
  $Q$ for every $a\in\Delta$ and $q\in Q$.
\end{definition}

A \emph{run} of $A$ on a tree $t$ is a labeling $\lambda:\dom(t) \to
Q$ such that for every $v\in\Dom(t)$ with $n$ children we have that
$\lambda(v1)\cdots \lambda(vn) \in\delta(\lambda(v),\lab^t(v)).$ Note
that when $v$ has no children, the criterion reduces to $\varepsilon
\in \delta(\lambda(v),\lab^t(v))$. A run is \emph{accepting} iff the
root is labeled with an accepting state, that is,
$\lambda(\varepsilon) \in F$. A tree is \emph{accepted} if there is an
accepting run. The set of all accepted trees is denoted by $L(A)$.

The regular languages encoding the transition function of an NTA are
represented by NFAs. The size of an NTA is then $|Q|+|\Delta|$ plus
the sizes of the NFAs.

A \defem{deterministic tree automaton (DTA)} is an NTA where
$\delta(q,a)\cap\delta(q',a)=\emptyset$ for all $a\in \Delta$ and
$q\neq q'\in Q$. The transition function of a DTA is represented by
DFAs.

The following is well known. 
\begin{lemma}\label{silly}

  \begin{enumerate}
  \item Deciding whether, for a given NTA $A$, $L(A)=\emptyset$ is in {\sc
      ptime}. \cite[Theorem 19]{martensneventcs}

  \item Testing whether, for given DTAs $A$ and $B$,  $L(A)\subseteq L(B)$
    is in {\sc ptime}. \cite[Theorem 3]{martensjoachim}

  \item Given a DTD $d$, a DTA $A_d$  such that $L(d)=L(A_d)$ can be
    constructed in exponential time. (essentially, \cite{bmw})
  \end{enumerate}
\end{lemma}


\section{Containment in the presence of disjunction}
\label{sec:disjunction}

Miklau and Suciu showed that \xcont for
XP(\child,\descendant,\filter,\wildcard) is {\sc
conp}-complete~\cite{miklaujacm}. In this section, we consider the
addition of disjunction to this fragment and show that \xcont remains
in {\sc conp}.  The problem remains hard for {\sc conp} even if only
the child or only the descendant axis is allowed together with
disjunction.  Miklau and Suciu already mention these results but do
not provide full proofs \cite{miklaujacm} (with exception of the lower
bound for XP(\child,\disjunction)). Therefore, we decided to include
the proofs in this paper.

We were surprised by the fact that the
{\sc conp} upper bound strongly depends on the fact that the alphabet
$\Sigma$ is infinite. Let \xfcont be the containment problem, where
additionally a finite set of labels is given as input and the
containment only has to hold for documents with labels from this set.
In Theorem~\ref{theo:finitelower}, we show that even for
 XP(\child,\descendant,\disjunction) the problem \xfcont is hard for
{\sc pspace}.

\subsection{Unrestricted alphabet}

\begin{theorem}\label{theo:disjunctionupper}
\xcont for $\XP(\child,\descendant,\filter,\wildcard,\disjunction)$
expressions is in  {\sc conp}.
\end{theorem}
\proof 
 We develop a criterion which allows to check in {\sc np}
    whether, for given expressions $p$ and $q$, $p\not\subseteq q$. Let
    $p$ and $q$ be fixed and let $ p_1 |\ldots | p_l$ and $q_1 | \ldots
    | q_{l'}$ be the disjunctive normal forms (DNFs) of $p$ and $q$,
    respectively.  Hence, each $p_i$ and $q_j$ is an expression from
    $\XP(\child,\descendant,\filter,\wildcard)$. Let $n$ and $m$
    denote the maximum number of nodes in an expression $p_i$ and $q_j$,
    respectively. Let $T(n,m)$ be the set of trees with at most
    $2n(m+2)$ nodes that are labeled with symbols that occur in $p$
    and with the new symbol $\#$ not occurring in $p$ nor in $q$. We
    prove the following claim:\\
  
  {\bf Claim.} $p \not\subseteq q \Leftrightarrow {}$ there is a $t\in
  T(n,m)$ such that $t\models p$ but $t\not\models q$.\\
  
  Clearly ``$\Leftarrow$'' holds. For the other direction we assume
  that there is a tree $s$ matching $p$ but not $q$. Then $s$ has to
  match one of the $p_i$. Hence, there is a homomorphism $f$ from
  $p_i$ to $s$.  i.e., $h$ maps the nodes of the tree pattern of $p_i$
  to the nodes of $s$ such that (1) $h(v)$ has the same label as $v$
  unless $v$ carries a wildcard, (2) $h(v)$ is a child (descendant) of
  $h(u)$ if and only if $v$ is a child (descendant) of $u$.

  We construct $t$ by transforming $s$ in several steps. Let $V$
  denote the set of nodes of $s$ in the image of $f$.  We delete all
  nodes in $s$ that are neither in $V$ nor an ancestor of a node in
  $V$.  The resulting tree, $t_1$, has at most as many leaves as
  $p_i$. We replace the labels of those nodes of $t_1$ which are not
  in $V$ by $\#$ and obtain $t_2$. Let $V'$ be the set of branching
  nodes of $t_2$, i.e., those nodes that have more than one child.
  The set $V'$ contains at most $n$ vertices.  Let a {\em pure path}
  of $t_2$ be a path without nodes from $V\cup V'$.  In particular,
  the nodes of a pure path are all labeled with $\#$.  We get $t$ by
  replacing in $t_2$ each maximal pure path with $>m+1$ inner nodes
  by a path with $m+1$ $\#$-labeled inner nodes. We refer to the
  nodes of $t$ which are inserted in this last step as {\em special
    nodes}. Clearly, there is a one-to-one correspondence between
  non-special nodes in $t_2$ and $t$. For a non-special node $u$ in
  $t$, the corresponding node in $t_2$ is denoted by $\tilde u$.

It is easy to see that $t\in T(n,m)$, that $t\models p_i$ and that $t$
contains at most $m+2$ times $|V|+|V'|$, hence $\le 2n(m+2)$, many nodes.

We have to show that $t\not\models q$. Towards a contradiction assume
that $t\models q_j$, for some $j$. Hence, there is a homomorphism
$h:q_j\to t$. Next, we show that $h$ can be modified to obtain a
homomorphism from $q_j$ to $s$ which leads to the desired
contradiction. We first define a homomorphism $h_2$ from $q_j$ to
$t_2$ as follows. Whenever $h(v)$ is a non-special node then
$h_2(v)=\widetilde{h(v)}$. 

Let $v_1,\ldots,v_k$ be nodes of $q_j$ such that
$h(v_1),\ldots,h(v_k)$ are special nodes which lie on some path of
$t_2$ which consists entirely  of special nodes,  ordered from the root
to the leaves. By the choice
of $m$ it holds that $k\le m$, therefore there must be an $i$ such
that $h(v_{i+1})$ is not a child of $h(v_i)$ or $h(v_1)$ is not the
first node of the path or $h(v_k)$ is not the last node of the path.

In either case, we can define $h_2$ for nodes from
$\{v_1,\ldots,v_k\}$ such that the child and descendant relations are
respected. In this way, we get a homomorphism $h_2$ from $q_j$ to $t_2$.

Clearly, all nodes of $q_j$ which are not mapped to nodes in $V$
must be labeled with a $\wildcard$. Thus, $h_2$ also defines a
homomorphism from $q_j$ to $t_1$ and to $s$, the desired contradiction.
This completes the proof of the claim.\\

It remains to show how the criterion of the above claim can be used for an {\sc
  np}-algorithm that checks whether $p\not\subseteq q$. The algorithm
  simply guesses an expression $p_i$ from the DNF of $p$ (by
  non-deterministically choosing one alternative for each $|$ in $p$) and
  a $t\in T(n,m)$. Then it checks that $t\models p_i$ and
  $t\not\models q$.  The latter can be done in polynomial time as
  shown in \cite{kochvldb}.\qed

\begin{theorem}\label{theo:disjunctionlower}
\begin{enumerate}[(a)]
\item \xcont for $\XP(\child,\disjunction)$ is {\sc conp}-hard.
\item \xcont for $\XP(\descendant,\disjunction)$ is
  {\sc conp}-hard. 
\end{enumerate}
\end{theorem}
\proof 
  \begin{enumerate}[(a)]
\item   The hardness proof is the same proof that shows that containment of
  regular expressions is {\sc conp}-hard~\cite{lewispapad}. We give it
  for completeness sake and because the next proof depends on it.We
  use a reduction from validity of propositional logic formulas in
  disjunctive normal form which is known to be complete for {\sc
    conp}~\cite{lewispapad}.  Let $\varphi=\bigvee_{i=1}^m C_i$ be a
  propositional formula in disjunctive normal form over the variables
  $x_1,\ldots,x_n$.  Here, each $C_i$ is a conjunction of literals.
  For a disjunct $C$ let $\tilde C$ be the expression
  $a_1/\cdots/a_n$ where
  $$a_i:=\left\{\begin{array}{ll}
   0&\text{if $\neg x_i$ occurs in $C$};\\
   1&\text{if $x_i$ occurs in $C$};\\
   (0|1)& \text{otherwise}.
\end{array}\right.$$ 
Let $\tilde q$ be the disjunction of the expressions
$\tilde C_i$, $i=1,\ldots,m$. Further, let $p$ be the expression
$(0|1)/\cdots/(0|1)$ where $(0|1)$ is repeated $n$ times.  Clearly, $p
\subseteq \tilde q$ iff $\varphi$ is valid.
\item The reduction is similar to the one above except that we define
  $\bar C$ as $a_1\descendant a_2\descendant\allowbreak\cdots \descendant a_n$, $\bar q$ as the
  disjunction of the expressions $\bar C_i$, $i=1,\ldots,m$, and $p$
  as $(0|1)\descendant\cdots\descendant(0|1)$.  We show that $p \subseteq \bar q
  \Leftrightarrow \varphi$ is valid. Suppose $p \subseteq \bar q$,
  then in particular $\bar q$ matches every 0-1-string of length $n$,
  hence, $\varphi$ is valid.  To prove the converse direction, suppose
  $\varphi$ is valid. If $p$ matches a branch in a tree then there
  are in particular $n$ positions with 0 or 1. The $i$-th such position
  can be seen as a truth assignment to $x_i$. As $\varphi$ is valid
  all possible assignments are accounted for by $\bar q$, and $\bar q$
  matches that branch.\qed
  \end{enumerate}

\subsection{Finite alphabet}

\label{sec:finite}

As mentioned above, when the alphabet is finite, and given as part of
the input, containment becomes much harder. In the rest of this
section, $\Sigma$ is therefore a \emph{finite} alphabet.

\begin{theorem}\label{theo:finitelower}
\xfcont for
  $\XP(\child,\descendant,\disjunction)$ is {\sc pspace}-hard.
\end{theorem}
\proof 
 We make use of a reduction from {\sc corridor
    tiling} which is known to be hard for {\sc
    pspace}~\cite{Chlebus86}.  Let $T=(D,H,V,\bar b, \bar t, n)$ be a
  tiling system.  Here, $D=\{a_1,\ldots,a_k\}$ is a finite set of
  tiles; $H,V\subseteq D\times D$ are horizontal and vertical
  constraints, respectively; $\bar b=(b_1,\ldots,b_n)$ and $\bar
  t=(t_1,\ldots,t_n)$ are $n$-tuples of tiles; and, $n$ is a natural
  number in unary notation. The question is whether there exists a
    number $m$ and a valid tiling of a board with $n$ columns and $m$
    rows. Here, a tiling is valid if the following conditions are
    fulfilled:
    \begin{itemize}
    \item The  bottom row is tiled with $\bar b$.
    \item The top row is tiled with $\bar t$.
    \item For each horizontal pair $(x,y)$ of tiles, $(x,y)\in H$.
    \item For each vertical pair $(x,y)$ of tiles ($y$ above $x$),
    $(x,y)\in V$. 
    \end{itemize}

  We use a string representation of the board where
  every row is delimited by \# and the last symbol is $\$$.  The
  expression $q$ selects all strings that do not encode a tiling. As
  $\Sigma$ we take $D\cup\{\#,\$\}$.  For
  $S=\{c_1,\ldots,c_n\}\subseteq \Sigma$, we abbreviate the expression
  $(c_1 \mid \cdots \mid c_n)$ by $S$. For an expression $r$, $r^i$
  denotes $r/\cdots/r$ with $i$ occurrences of $r$.
  The expression $p$ is $.\descendant\$$
  assuring that
  the string contains the symbol $\$$. The expression $q$ is the
  disjunction of the following expressions. 
  \begin{itemize}
  \item some row has the wrong format: 
    \begin{itemize}
    \item some inner row has too few tiles: $\bigcup_{i=0}^{n-1} .\descendant\#D^i\#$

    \item the first row has too few tiles: $\bigcup_{i=0}^{n-1} D^i/\# $
      
    \item the last row has not enough tiles: $ \bigcup_{i=0}^{n-1} .\descendant\#D^i/\$$

    \item some row has too many tiles:  $.\descendant D^{n+1};$

    \end{itemize}

  \item $\$$
    occurs {\em inside} the string: $.\descendant\$/(D\cup \{\$\}\cup\{\#\})$;

  \item the string does not begin with $\bar b$: $\bigcup_{i=1}^n
    b_1/\cdots/b_{i-1}/(\bigcup_{a_j\not=b_i} a_j)$;
  \item the string does not end with $\bar t$: $\bigcup_{i=1}^n
    .\descendant (\bigcup_{a_j\not=t_i} a_j)/t_{i+1}/\cdots/t_n/\$$
  \item some vertical constraint is violated:$
    \bigcup_{(d_1,d_2)\not \in V} .\descendant d_1/(D\cup\{\#\})^{n}/d_2$; and,
  \item some horizontal constraint is violated: $\bigcup_{(d_1,d_2)\not
      \in H} .\descendant d_1/d_2$.  
\end{itemize} 

Now, $T$ has a solution iff $p\not\subseteq q$.  Clearly, if $T$ has a
solution then we can take the string encoding of the tiling as a
counter example for the containment of $p$ and $q$.  Conversely, if
$p\not\subseteq q$ then there is a, not necessarily unary, tree $t$
with one branch $s$ ending on a $\$$ such that $s\models p$ and $s\not\models
q$.  So, this branch encodes a solution for $T$.\qed

Actually, in the proof of Theorem \ref{theo:finitelower}, the
restriction to a finite alphabet is only used to express that a
certain element name in the XML document does not occur in a certain
set.  Therefore, if we extended the formalism with an operator
$*_{\not\in S}$ for a finite set $S$, expressing that any symbol but
one from $S$ is allowed, then containment would also be hard for {\sc
  pspace}.

\bigskip

For the upper bound we need the notion of alternating tree
automata~\cite{slutzkialternating} which is defined next. These
automata operate on trees where every node has rank at most $k$ (for
some fixed $k$). That is, every node has at most $k$ children.

\begin{definition}
  \rm An \defem{alternating tree automaton (ATA)}
  is a tuple $A=(k,Q,\Sigma,q_0,\delta)$ where $k>0$, $Q$ is a finite
  set of states, $\Sigma$ is the finite alphabet, $q_0\in Q$ is the
  initial state, and $\delta:Q\times\Sigma\times\{0,1,\ldots,k\}\to
  {\bf B}^+(\{0,1,\ldots,k\}\times Q)$ is the transition function.
  Here, ${\bf B}^+(\{1,\ldots,k\}\times Q)$ denotes the set of
  positive Boolean formulas over the set $\{1,\ldots,k\}\times Q$.
  
  A configuration on a tree $t$ is a tuple $[u,q]$ where $u\in\dom(t)$
  and $q\in Q$. An \defem{accepting run} of $A$ on $t$ is a tree $s$
  where nodes are labeled with configurations such that the root of
  $s$ is labeled with $[\epsilon, q_0]$, where $\epsilon$ is the root
  of $t$ and, for every node $u$ of $s$ (including leaf nodes), the
  following local consistency condition holds.  Let $u$ be labeled
  with $[v,q]$ with $n$ children labeled $[v_1,q_1],\ldots,
  [v_n,q_n]$. Then it must hold that
    \begin{itemize}
    \item each $v_i$ is a child of $v$ or $v$ itself; and,
    \item $\delta(q,\lab_t(v),m)$ is satisfied by the truth assignment
      $\rho$, where $m\leq k$ is the number of children of $v$ in $t$,
      and $\rho((\ell,q'))$ is true if for some $i$, $q_i=q'$ and
      $v_i$ is the $\ell$-th child of $v$ (where we view $v$ itself as
      the $0$-th child).
    \end{itemize}
  A tree is \defem{accepted} by $A$ if there is an accepting run. By $L(A)$
we denote the set of trees accepted by $A$.
\end{definition}

Note that ATAs as we defined them do not have final states. These are
encoded by transitions of the form $\delta(q,\sigma,0)=\text{\tt
  true}$.

\begin{theorem}\label{theo:finiteupper}
\xfcont for
  $\XP(\child,\descendant,\filter,\wildcard,\disjunction)$
  is in {\sc pspace}.
\end{theorem}
\proof 
  We show first that $p\not\subs q$ implies that there is a counter
  example tree with small degree and only a few  branching
  nodes.  More precisely, we call a tree \defem{$k$-bounded} if it has
  at most $k$ non-unary nodes (that is, nodes with more than one
  child) and every node has rank at most $k$.  For an
  $\XP(\child,\descendant,\filter,\wildcard,\disjunction)$ -expression
  $p$ let $f(p)$ be the maximum number of filter expressions in any
  disjunct of the DNF of $p$. We claim that $p\subs q$ if and only if
  $t\models p$ implies $t\models q$ on the class of $f(p)$-bounded
  trees.
  
  Indeed, suppose there is a $t$ such that $t\models p$ and $t\not
  \models q$. Let the DNF of $p$ and $q$ be $p_1 | \cdots | p_n$ and
  $q_1 | \cdots | q_m$, respectively.  Thus, for some $i$, $t\models
  p_i$, but $t\not\models q_j$, for all $j$. Let $h$ be a homomorphism
  from $p_i$ to $t$ and let $s$ be the tree obtained from $t$ by
  deleting all nodes that are neither in the image of $h$ nor
  ancestors of such nodes.  Clearly, $s$ is $f(p)$-bounded, $s\models
  p_i$ and $s\not\models q_j$ for all
  $j$ (otherwise, $t\models q_j$).\\

Next, we show that   for every $\XP(\child,\descendant,\filter,\wildcard,\disjunction)$
  expression $p$ there is an ATA $A_p$ such that for every
  $f(p)$-bounded tree $t$, $t\models p$ iff
  $A_p$ accepts $t$.  Moreover, $A_p$
  can be constructed in {\sc logspace}.

  To this end, let $p$ be an
  $\XP(\child,\descendant,\filter,\wildcard,\disjunction)$ expression.
  As the alphabet is finite and fixed, we can replace every
  $\wildcard$ with a disjunction of the alphabet symbols. Hence, we
  assume $p$ does not contain $\wildcard$.  Let $k=f(p)$. We define
  $A_p=(k,Q,\Sigma,q_0,\delta)$ where $q_0=p$ and $Q$ is the set of
  sub-expressions of $p$, all filter expressions of $p$ and all
  node tests of $p$.  Intuitively, a pair $[v,q]$ in an accepting run
  of $A_p$ on a tree $t$ means that $q$ holds in the sub-tree of $t$
  rooted at $v$.  For all $m\le k$, the transition function is
  inductively defined as follows:
  

  \begin{itemize}
\item $\delta({\tself :: \sigma [e_1]\cdots[e_\ell]}/p',\sigma,m)=
 (0,p') \wedge \bigwedge_{i=1}^\ell (0,e_i)$;

 \item $\delta(\tchild :: \sigma [e_1]\cdots[e_\ell]/p',\tau,m)=
 \bigvee_{j=1}^m  [(j,\sigma) \wedge (j,p') \wedge \bigwedge_{i=1}^\ell (j,e_i)]$;
 \item $\delta(\tdescendant :: \sigma
     [e_1]\cdots[e_\ell]/p',\tau,m)=$\\
 \mbox{ } \hfill
 $\bigvee_{j=1}^m (j,\semt{\tdescendant :: \sigma
     [e_1]\cdots[e_\ell]/p'} \vee
 ((j,\sigma) \wedge (j,p') \wedge \bigwedge_{i=1}^\ell (j,e_i))$;

\item $\delta(\sigma,\sigma,m)=\mytrue$;
\item $\delta(p \mid q,\sigma,m)= (0,p) \vee (0,q)$; and,
\item $\delta(e_1 \mid e_2,\sigma,m)= (0,e_1) \vee (0,e_2)$.
\end{itemize}

  The combinations $\delta(p,\sigma,m)$ that are not mentioned are
  \myfalse.  
For location paths of length 1, i.e., if in one of the first three
  transitions there is no $p'$, the atoms $(0,p')$ or $(j,p')$ are
  removed, respectively. 
It is straightforward to prove by a nested induction on the structure
  of the tree and the expression that the pairs $[v,q]$
have the intended meaning. Therefore, a tree $t$ has an accepting run
of $A_p$ if and only if $t\models p$.\\

Thus, to decide $p\subs q$ it is sufficient to test whether every
$f(p)$-bounded tree accepted by $A_p$ is also accepted by $A_q$. Note
that $f(p)$-bounded trees can be easily encoded by strings.  We say
that a node is a fork if it has more than one child. As there are at
most $f(p)$ forks, every tree consists of at most $k:=f(p)\times f(p)$ unary
paths that are joined at the at most $f(p)$ forks. To every path we
associate its lower fork (or none if there is no such fork).  Next, we assign a
unique number to each path such that higher paths get lower numbers.
Let $s_i$ be the
concatenation of the labels on path $i$ and let $i_1,\ldots, i_\ell$
be the paths rooted at the fork below branch $i$. Let
$a_i:=s_i\, i_1\,\cdots\, i_\ell\#$.
Every $f(p)$-bounded tree $t$ can then be encoded by the string
$a_1\cdots a_m$. Let $A'_p$ and $A'_q$ be the alternating string
automata that simulate $A_p$ and $A_q$ on the string representations
of bounded trees. Basically, whenever the automaton reaches the end of
(the encoding of) a path, the numbers $i_1,\ldots,i_\ell$ indicate the
positions of the children paths and it can reach a path $i_j$ by
skipping everything before its occurrence.
This is possible as all paths are ordered and higher
numbered paths occur to the right.  Finally, let $A$ be the
automaton that checks whether the input string is a valid encoding of
a bounded tree.  The problem then reduces to testing whether $A \cap
A'_p \cap \neg A'_q$ is empty. The latter can be done in {\sc
  pspace}~\cite{alternationCKS}. We arrive at the desired result.\qed


\newcommand {\alga}{{\tt CheckPnotq}\xspace} 
\newcommand {\algb}{{\tt CheckP}\xspace} 
\newcommand {\algc}{{\tt Checknotq}\xspace}

\section{Containment in the presence of DTDs}
\label{sec:dtd}

In this section we study the \xdcont problem, i.e., the containment
problem relative to a DTD. Deutsch and Tannen \cite{dtxpath} show a
{\sc pspace} lower bound for \xdcont for \XPfull.  Marx
\cite{marxedbt} gives an  {\sc exptime} upper bound for a much larger
fragment, including all axes and negation in filter expressions.

We show here that \xdcont for \XPfull
problem is actually {\sc exptime}-complete.  We also exhibit a simple
fragment with tractable \xdcont and a modest {\sc np}-completeness
result on the fragment using only 
\child\ and \filter. We do not know how to extend the upper bound
proof to include \descendant\ or \wildcard.   The results of this
section are summarized in Table~\ref{tab:dtd}.

We illustrate by a simple example that the presence of a DTD can
complicate matters. Consider the DTD
\begin{eqnarray*}
  a & \to & ab \mid \epsilon\\
  b & \to & c\\
  c & \to & \epsilon
\end{eqnarray*}
and the expressions $p=a/a$ and $q=.//b/c$. Although $p$ and $q$ are
seemingly unrelated and, in particular, it is not the case that every
path matching $p$ also matches $q$, it holds that each tree which
respects the DTD and matches $p$ also matches $q$.

We remark that it can be tested in polynomial time whether, for a DTD
$d$ and a symbol $a$, there exists a tree $t\in L(d)$ with a vertex
labeled $a$. Therefore, we assume in the following that each input
DTD contains only {\em useful} symbols. In particular, for each $d$
and each symbol $a\in\Sigma_d$, there is a tree valid with respect to
$(d,a,\Sigma_d)$. 

\subsection{A tractable fragment}

We start with a fragment in {\sc p}.

\begin{theorem}\label{theo:dtd:p}
\xdcont of $\XP(DTD,\child,\descendant)$-expressions is in {\sc P}.
\end{theorem}
\proof 
  Let $d$ be a DTD and $p,q$ be expressions of
  $\XP(DTD,\child,\descendant)$.
  
  We first show how to construct a non-deterministic top-down
  automaton $A_p$ which checks that, for a tree $t$, $t\models p$
  holds.  To this end, let $p=p_1//p_2//\cdots//p_k$, where in each
  $p_i$ only the child axis is used. For each $i$, let $p_i$ contain
  $i_\ell$ child-axis location steps.
  
  Intuitively, $A_p$ guesses a path in $t$ which matches $p$ and, for
  each node $v$ on this path it maintains the lexicographically
  maximal $(i,j)$ such that  the path from the root to $v$ matches the
  expression $p_1//\cdots//p_{i-1}//p_i^j$, where $p_i^j$ consists of the first
    $j$ location steps of $p_i$ (along the child-axis).

  It should be stressed here, that $A_p$ needs non-determinism only to
  guess the path. The computation of the pairs $(i,j)$ is completely
  deterministic and similar to the case of the standard string pattern
  matching automata~\cite{dataalgbible}. The automaton, $A_p$ enters
  an accepting state on the leaf $u$ of the distinguished (guessed)
  path if and only if the computed pair for $u$ is $(k,i_k)$. On all
  other leaves it takes an accepting state in any case.

  An automaton $A_q$ which accepts all trees that do {\em not} match
  $q$ can be constructed along the same lines. It computes a pair
  $(i,j)$ with the same intended meaning as above, for every node $v$
  of the tree and takes an accepting state at all leaves that have
  {\em not} reached $(k,i_k)$. This automaton is actually
  deterministic.
 
  By combining $A_p$ with $A_q$ and the canonical non-deterministic
  automaton $A_d$ which tests $t\models d$, we obtain an automaton
  which accepts all counterexamples to $p\subs_d q$. As this automaton
  is of polynomial size in $p,q,d$ and testing emptiness of
  non-deterministic tree automata is in {\sc ptime}
  (Lemma~\ref{silly}(1)), we obtain the stated upper bound.\qed

It should be mentioned that in \cite{NevenSXPath03} we claimed that
\xdcont of $\XP(DTD,\child,\descendant,$ $\wildcard)$-expressions is in
{\sc p}. Unfortunately, we were not able to extend the proof sketch
given there into a complete proof. In fact, we conjecture that this
problem is {\sc conp}-hard.

\subsection{Fragments in {\sc conp}}

Next, we consider a fragment in {\sc conp}. It is open whether
$\XP(DTD, \child, \filter)$ is a maximal fragment whose complexity of
containment w.r.t.~DTDs is in {\sc conp}.

\begin{theorem}\label{theo:dtd:npupper}
\xdcont for $\XP(DTD, \child, \filter)$ is in {\sc
    conp}.
\end{theorem}
\proof 
The obvious idea is to guess a tree $t$ which  matches $P$ but not $q$.
A complication arises from the fact that the smallest such
tree $t$ might be of exponential size due to the constraints from
$d$. Thus, we give a non-deterministic algorithm \alga$(d,a,P,q)$ which checks,
  given a DTD $d$, a non-terminal $a$ of $d$, a set
  $P=\{p_1,\ldots,p_n\}$ of $\XP(\child, \filter)$-expressions, and an
  $\XP(\child, \filter)$-expression $q$, whether there is a tree $t$
  with root symbol $a$ which conforms to $d$, matches all expressions in
  $P$ but does not match $q$. But it does not explicitly construct
  such a tree.  Clearly, invoking this algorithm with
  $d$, $q$, $P=\{p\}$ and $a$ as the start symbol of $d$ checks
  whether $p\not\subseteq_d q$. We note that we allow a set $P$ of
  expressions as input for \alga because the algorithm uses such sets
  for recursive calls.

Algorithm \alga makes use of two algorithms with slightly simpler tasks. Algorithm
{\algb} checks on input $d,s,P$ whether there is a tree $t$ with root
$s$ conforming to $d$ which contains all the expressions from $P$.
Algorithm $\algc$ checks on input $d,q$ whether there is a tree
conforming to $d$ with a root labelled by the root symbol of $q$ which
does {\em not} match $q$.

We assume in the following that all labels of $P$ and $q$ occur in $d$
and that $d$ only contains symbols from which a tree can be
derived. By $s(p)$ we denote the root symbol of an expression $p$, i.e., the
label of the root of $\tau(p)$. A level 1 sub-expression of an
expression $p$ is an expression corresponding to a child of the root
in $\tau(p)$. Let $l$ denote the overall number of
  depth-1-nodes in expressions of $P$.

\NC{\qval}{\text{qval}\xspace}
\begin{figure}
  \centering
\algo{
$\algc(d,a,q)$\\
(Returns TRUE if there exists a tree with root $a$ which does not match $q$)
\begin{enumerate}[1.]
 \item If $s(q)\not=a$ return TRUE.
\item If $\tau(q)$ has only one node return FALSE.
\item Guess a string $u\in d(a)$ of length $\le |d|$ and a level 1
  sub-expression $q'$ of $q$. 
\item If $b:=s(q')$ does not occur in $u$ return TRUE.
\item Return $\algc(d,b,q')$.
\end{enumerate}
}

  \algo{
$\algb(d,a,P=\{p_1,\ldots,p_n\})$\\
(Returns TRUE if there exists a tree with root $a$ matching all $p_i$)
\begin{enumerate}[1.]
\item If some expression in $P$ does not have the root
symbol $a$  return FALSE.
\item Guess a string $u\in d(a)$ of length $\le(|d|+1)(l+2)$.
\item For each $i\in\{1,\ldots,n\}$, guess a mapping $f_i$ from
  the level 1 sub-expressions of $p_i$ to the
  positions of $u$.
\item For each position $j$ of $u$, which is in the image of at least
  one of the mappings $f_i$
  \begin{enumerate}[(a)]
  \item Let $P'$ be the set of level 1 sub-expressions $p$ with $f_i(p)=j$.
  \item Call $\algb(d,u_j,P')$.
  \end{enumerate}
\item Return TRUE iff all the recursive calls return TRUE.
\end{enumerate}
}  

\algo{
$\alga(d,a,P=\{p_1,\ldots,p_n\},q)$\\
(Returns TRUE if there exists a tree with root $a$ matching all $p_i$ but not $q$)
\begin{enumerate}[1.]
\item If some expression in $P$ does not have the root
symbol $a$ THEN return FALSE.
\item If $s(q)\not=a$ return $\algb(d,a,P)$.
\item If $\tau(q)$ has only one node return FALSE.
\item Guess a level 1 sub-expression $q'$ of $q$.
\item Guess a string $u\in d(a)$ of length $\le(|d|+1)(l+2)$.
 \item If $b:=s(q')$ occurs in $u$ call $\alga(d,b,\emptyset,q')$. 
\item For each $i\in\{1,\ldots,n\}$, guess a mapping $f_i$ from
  the level 1 sub-expressions of $p_i$ to the
  positions of $u$.
\item For each position $j$ of $u$, which is in the image of at least
  one of the mappings $f_i$
  \begin{enumerate}[(a)]
  \item Let $P'$ be the set of level 1 sub-expressions $p$ with $f_i(p)=j$.
  \item Call $\alga(d,u_j,P',q')$.
  \end{enumerate}
\item Return TRUE iff all the recursive calls return TRUE.
\end{enumerate}
}  
  \caption{Algorithms \algc, \algb and \alga used in the proof of
    Theorem \ref{theo:dtd:npupper}.}
  \label{alg:alga}
\end{figure}

The algorithms are given in Figure \ref{alg:alga}.   The
algorithms follow a top-down approach and work recursively.  The
correctness can be shown by induction on the number of 
recursive calls. \algc and \algb are quite straightforward.

An important point is that in \algb it is sufficient to consider strings $u$ of
 length at most $(|d|+1)(l+2)$. It can be shown by a simple pumping
 argument that if a tree matching $P$ has a level with more children
 then there is a sub-sequence of these children which is not in the
 image of any mapping and can be removed without leaving $d(a)$.

\alga is basically a combination of \algc and \algb. It has to check
that there is a path in $q$ which does not match any path in the
counter-example tree. Step 6 is needed to verify that this also holds
in those parts of the tree which are not needed to fulfil $P$.

It remains to show that this (non-deterministic) algorithm works in
polynomial time. This follows directly from the fact that for each
node in $\tau(p)$ and each node $v$ in $\tau(q)$ there is at most one
recursive call of $\alga$ in which $v$ is the root of some expression in
$P$ (or $q'$).\qed

The next theorem follows directly from \cite{woodwebdb}.

\begin{theorem}\label{theo:dtd:nplower}
\begin{enumerate}[(a)]
\item
\xdcont for $\XP(\child,\filter)$
is {\sc conp}-hard.
\item
\xdcont for $\XP(\descendant,\filter)$
is {\sc conp}-hard.
\end{enumerate}
\end{theorem}

\proof 
In \cite{woodwebdb}, the following problem is shown
to be {\sc conp}-hard.

\smallskip

{\sc sibling constraint implication (sc imp)}:\\
Given: Regular expression $r$ over alphabet $\Sigma$, a set
$S\subseteq\Sigma$, and $a\in \Sigma$.\\
Question: Does every string $w\in L(r)$ that contains all the symbols
in $S$ also contain the symbol $a$ (denoted $r\models S\to a$)?
 
\smallskip

We reduce {\sc sc imp} to $\XP(\child,\filter)$ and
$\XP(\descendant,\filter)$ . Thereto, assume given $r$,
$S=\{s_1,\ldots,s_k\}$, and $a$. Construct the DTD $d$ consisting of
the sole rule $\text{start}\to r$, where $\text{start}\not \in
\Sigma$, then $\text{start}[s_1]\cdots [s_k] \subseteq_d
\text{start}[a]$ iff $\text{start}[.//s_1]\cdots [.//s_k]
\subseteq_d \text{start}[.//a]$ iff $r\models S\to a$.\qed

\subsection{Fragments in {\sc exptime}}

When both the child and descendant axes are allowed, then adding
filter expressions and wild-card or disjunction raises the complexity
of \xdcont to {\sc exptime}.

Although in \cite{marxedbt} it is shown that \xdcont is in {\sc
  exptime} even for
full navigational XPath by a reduction to propositional
dynamic logic, we give here a simpler proof of the result for our
  downwards navigating fragment. 

First, we introduce two concepts. Let $p$ be an XPath-expression.  An
expression is a \defem{sub-expression} of an expression $p$ if it is
generated by an lpath node in the derivation tree of $p$ according to
the grammar in Definition~\ref{def:syn}.

The \defem{\tself-closure} of $p$, denoted by $\tself(p)$, is
inductively defined as follows: for a location step
$p_1=\text{axis}::\sigma[e_1]\cdots[e_k]$ its \tself-closure, is
$\tself::\sigma[e_1]\cdots[e_k]$. For expressions $p_1/p_2$ and
$p_1\mid p_2$, their \tself-closure is $\tself(p_1)/p_2$ and
$\tself(p_1)\mid \tself(p_2)$, respectively.
\begin{theorem}\label{theo:dtd:exptimeupper}
\xdcont for $\XP(\child,\descendant,
\filter,\wildcard,\disjunction)$
is in {\sc exptime}.
\end{theorem}
\proof 

  We provide a translation to
  containment of unranked deterministic tree automata whose size is exponential in
  the input. By Lemma~\ref{silly}(2), the latter is in {\sc exptime}.
  
  We first show that for each $\XP(\child,\descendant,\allowbreak
  \filter, \allowbreak \wildcard, \disjunction)$-expression $p$, one
  can construct in exponential time an exponential size deterministic
  tree automaton $A_p=(Q,\Sigma_d,\delta,F)$ such that $A_p$ accepts a
  tree $t$ if and only if $t\models p$.  Here, $\Sigma_d$ is the
  finite alphabet associated to the given DTD $d$.  The states of
  $A_p$ are pairs $(S,D)$ where $S$ and $D$ are sets of sub-expressions
  of $p$ or the \tself-closure of sub-expressions of $p$.

  The intended meaning of the states is as follows. If $p_1\in S$ at
  some vertex $v$ of a tree $t$ then $t_v\models p_1$.  If $p_1\in D$
  then there is some node $w$ below $v$ in $t$ such that $t_w\models
  p_1$.  So, $S$ describes all expressions that hold at the current node,
  while $D$ describes all expressions that hold at descendants of the
  current node.  

  Set $F=\{(S,D)\mid p\in S\}$.  It remains to define the transition
  function for each $\delta((S,D),a)$. Recall that the corresponding
  DFA operates on strings of the form: $(S_1,D_1)\cdots
  (S_\ell,D_\ell)$. Then $S$ should contain exactly the
  XPath-expressions generated by the following rules.
  \begin{itemize}
  \item $\tchild::\sigma[e_1]\cdots[e_k]/p'\in S$ iff there is an $i\le\ell$
   such that $\tself::\sigma[e_1]\cdots[e_k]/p'\in S_i$;
 \item $\tdescendant::\sigma[e_1]\cdots[e_k]/p'\in S$ iff there is an $i\le\ell$
   such that $\tself::\sigma[e_1]\cdots[e_k]/p'\in S_i\cup D_i$;

  \item $\tself::\sigma[e_1]\cdots[e_k]/p'\in S$ iff $\sigma=a$ or
    $\sigma=*$, $p'\in S$ and $e_i\in S$ for $i=1,\ldots,k$;

  \item $p_1\mid p_2 \in S$ iff $p_1\in S$ or $p_2\in S$; and,

\item $p_1 \text{ or } p_2 \in S$ iff $p_1\in S$ or $p_2\in S$.
  \end{itemize}
For the case without a sub-expression $p'$ the first three rules are
adapted in the obvious way.

For each expression $p$, $p\in D$ if $p\in S_i\cup D_i$, for some $i$.

  It remains to describe how a DFA $B$ of exponential size can execute
  the above rules. There is a linear number of sub-expressions of $p$
  and self-closures of those.  The DFA $B$ keeps for each of them one
  bit in memory indicating whether the corresponding expression is in
  $S$ or $D$. Initially none of them are.  An expression is put in
  a set if one of the above rules fire.  Every rule should be checked
  at every transition step.  So, the size of each $B$  is exponential
  in $p$. As $A_p$ contains an exponential number of such DFAs, its
  size is also exponential.

Let $A_d$ be the exponential size deterministic automaton accepting
$d$ (cf.~Lemma~\ref{silly}(3)). Then deciding whether $p\subs_d q$
reduces to testing whether $L(A_d)\cap L(A_p)\subs L(A_q)$. By
Lemma~\ref{silly}(2), the latter can be done in {\sc exptime}.\qed 


\begin{theorem}\label{theo:dtd:exptimelower}
\xdcont for $\XP(\child,\descendant,
\disjunction)$
is hard for {\sc exptime}.
\end{theorem}
\proof 
  The proof makes use of a reduction from {\sc two-player corridor
    tiling}.  This is the extension of {\sc corridor tiling}, used in
  the proof of Theorem~\ref{theo:finitelower}, to two players.
  Let $T=(D,H,V,\bar b, \bar t, n)$ be a tiling system. Again, $D$ is
  a finite set of tiles; $H,V\subseteq D\times D$ are horizontal and
  vertical constraints, respectively; $\bar b$ and $\bar t$ are
  $n$-tuples of tiles; and $n$ is a natural number. There are two
  players (\I\ and \II) that place tiles in turn on an
  $n\times\Bbb{N}$ board.  On this board the bottom row is tiled with
  $\bar b$.  Player \I\ starts on the first square of the second row
    from the bottom.
  Each player in turn places a tile on the next free square going from
  left to right and from bottom to top. While player \I\ tries to
  construct a corridor tiling from $\bar b$ to $\bar t$, player \II\ 
  tries to prevent it.  If player \II places a tile which is not
  consistent with respect to the horizontal and vertical constraints
  then player \I can answer with a special tile `!'.  Player \I\ wins
  if a tiling is constructed satisfying the horizontal and vertical
  constraints with the top row tiled with $\bar t$, or if she answers
  an inconsistent tile placed by \II with `!'. We say that player \I
  has a winning strategy if she can always win no matter how \II\ 
  plays. It is well-known that it is {\sc exptime}-complete to
  determine whether \I\ has a winning strategy~\cite{Chlebus86}. This
  result even holds if the number of tiles in a row is forced to be
  even. Thus we assume in the following that $n$ is even.

  We encode strategies for player \I as trees. For each position in
  the game in which \II moves, the tree contains all possible moves of
  player \II and, for each \I-position, it contains only one
  move. Thus, such a tree encodes a winning strategy if and only if
  each path corresponds to a correct tiling or to a wrong move of
  \II. 

To this end, we use
  symbols of the form $(a,i)$, where $a\in D$ and $i\in\{1,2\}$ plus
  some additional auxiliary symbols: $\$$ to indicate the borders
  between rows, $\#$ to mark the end of a tiling and the ``protest
  symbol'' `!'. Inner nodes of the tree correspond either to moves of
  \I or \II. Nodes corresponding to \I are labeled $(a,1)$ where $a$
  is the tile chosen by \I in this move. They have one child for every
  tile corresponding to the possible next moves of \II. Nodes
  corresponding to \II are labeled $(a,2)$ and have only one child
  which is the unique answer move of \I according to her strategy or
  one of $\$$, $\#$ or `!'. Nodes with label $\$$ have one child
  corresponding to a move of \I. Nodes with label $\#$ or `!' are
  leaves.  The root of the tree is labeled $S$ and represents an
  empty (dummy) move which has to be answered by \I. It has one
  child. On each path from the root, \I-nodes and \II-nodes alternate
  (when we ignore the intermediate $\$$ nodes).

  Now we describe the reduction in more detail.  Let
  $D=\{d_1,\ldots,d_m\}$.  We use the following DTD $f$ which defines
  all possible strategy trees for player \I:
$$\begin{array}{l}
S \to (d_1,1) + \cdots + (d_m,1)\\
\$ \to (d_1,1) + \cdots + (d_m,1)\\
\text{and for every $d\in D$, we have the rules}\\
(d,1) \to (d_1,2) \cdots (d_m,2)\\
(d,2) \to (d_1,1) + \cdots + (d_m,1) + \# + \$ + !\\
! \to \epsilon\\
\# \to \epsilon
\end{array}
$$
Note that $\#$ and `!' are the only terminal symbols. Thus, each path
in the tree either ends with $\#$ or `!'. 

A derivation tree encodes a strategy tree (or game
tree) for \I. As
the bottom and the top row are fixed we do not represent them in these
trees, i.e., only intermediate rows are represented. We 
assume that the tiling consisting only of the top and bottom row is
not valid. Therefore any strategy tree has to represent at least one
row.

We have to check whether there is a tree encoding a valid strategy
tree for \I.  We will construct an expression $q$ which selects a tree
if and only if it does {\em not} encode a winning strategy for \I.
Thus, $S \subseteq_f q$ if and only if player \I\ has no winning
strategy.

We define $$q := q_{\text{$n$ tiles}}\mid q_V \mid q_H  \mid q_!,$$
where the sub-expressions on
the right hand side will be defined shortly.  Intuitively,
$q_{\text{$n$ tiles}}$ expresses that some row has a wrong length
(which can only be due to player \I), $q_V$ and $q_H$ express that
some vertical or horizontal constraint, respectively, is violated by
\I, and $q_!$ expresses that \I used the protest symbol wrongly. 
Each of these 
sub-expressions identifies an error in the strategy tree. Hence, if every
tree matches one of these expressions, every tree contains an error
and no tree can be a valid strategy tree.

Note that, although the expressions under consideration do not have
the wild-card available, the disjunction of all alphabet symbols
defined by the grammar is a kind of wild-card as the DTD assures that
no other symbols occur in the tree. We denote the set of all pairs
$(\sigma,i)$, where $\sigma$ is a tile and $i\in \{1,2\}$ by $\Sigma$.
Further, we denote by $\Sigma_\$$ the set $\Sigma\cup\{\$\}.$

Three types of errors can occur in a strategy tree: (1) the tree is of
the wrong shape; (2) player \I\ places an inconsistent
tile, or (3) player \I uses the symbol `!' although \II placed a
correct tile. 
\begin{description}
\item[A row does not contain exactly $n$ tiles.]
$$q_{\text{$n$ tiles}}:=
.\descendant D^{n+1} \mid \bigcup_{i=0}^{n-1}(\$ \mid S)/ 
D^{i}/(\$ \mid \#).
$$
Recall that we use $D$ as a shortcut for $d_1\mid\cdots\mid d_m$.

\item[Vertical Constraints are violated.] 
\[
q_V :=  q_{\bar b} \mid q_{\bar t} \mid \bigcup_{(d',d)\not \in V}
.\descendant(d',1)/\Sigma^{n}_\$/(d,1), 
\]
where 
$$q_{\bar b}:=\bigcup_{i=1}^n   
\bigcup_{(b_i,d)\not\in V}
S/\Sigma^{i-1}/(d,1)$$
    checks the vertical constraints w.r.t.{} $\bar
    b$, and
$$q_{\bar t}:=    
\bigcup_{i=1}^n\bigcup_{(d,t_i)\not\in V} .\descendant(d,1)/\Sigma^{n-i}/\#,$$
checks the vertical constraints w.r.t.{} $\bar t$.

\item[Horizontal Constraints are violated.]
\[
q_H := \bigcup_{(d',d)\not \in H} .\descendant(d',2)/(d,1).
\]

\item[Wrong use of `!'.] 
\[
q_! := (\bigcup_{(d_1,d)\in V\atop (d_2,d)\in H}
.\descendant(d_1,2)/\Sigma^n_\$/(d_2,1)/(d,2)/!) 
\;\cup\;
(\bigcup_{i=2}^{n}\bigcup_{(b_i,d)\in V\atop (d',d)\in H}
S/\Sigma^{i-2}/(d',1)/(d,2)/!)
\] 
This expression takes care of the case that player \I uses the protest
symbol `!' although the last tile placed by player \II was consistent
with the vertical and horizontal constraints. The second disjunct in
this expression takes care of the first row of the game (i.e., the
second row in the tiling).\qed

\end{description}


By similar techniques, making use of the techniques of Lemma 3
in~\cite{miklaujacm}, it can be shown that 
\xdcont for $\XP(\child,\descendant,\filter,\wildcard)$
is hard for {\sc exptime}.

The outermost union can be handled as in Lemma 3 of~\cite{miklaujacm}.
Of course, the DTD has to be adapted accordingly.


\section{Containment in the presence of variables}
\label{sec:variables}

\NC{\loctrue}{T\xspace}
\NC{\locfalse}{F\xspace}
\NC{\locor}{\text{ or }}

In this section, we study Boolean containment of XPath
expressions which allow the comparison of attribute values. More
precisely, in this section we consider XPath expression generated by
the following grammar.

\begin{center}
\begin{tabular}{rcl}
lpath & $::=$&  lstep $\mid$ lpath '/'
lpath $\mid$ lpath '$|$' lpath\\
lstep & $::=$ & axis '::' nodetest ('['expr']')$^*$\\
axis & $::=$ & \tself $\mid$ \tchild $\mid$ \tdescendant\\
expr & $::=$ & fexpr $\mid$ vexpr $\mid$ expr 'or' expr\\
fexpr & $::=$ & lpath\\
vexpr & $::=$ & variable $=$ attribute $\mid$ variable $\not=$ attribute\\
\end{tabular}
\end{center}

A variable $x$ is denoted as $\$x$, an attribute $a$ as $@a$.  In this
section, fragments without filter expressions still allow
sub-expressions of the type vexpr. Furthermore, fragments without
filter but with disjunction allow disjunctions of sub-expressions of
type vexpr (cf. Theorem~\ref{theo:evars:lower}(c)).

We consider two different semantics. The original XPath semantics and
the existential semantics, introduced in \cite{dtxpath}.

In the \defem{XPath semantics}, variable bindings are defined in an
outer context.  In
particular, the value of an expression is defined with respect to a
variable assignment $\rho:X\to \data$ where $X$ is the set of all
variables. We denote the truth value of a filter expression $e$
relative to a variable assignment $\rho$ and a vertex $v$ by
$E_t^\rho(v,e)$ and the semantics of an XPath expression $p$ by
$\semt{p}^\rho$.

Formally, we extend Definition~\ref{def:sem} as follows. Given a tree
$t$, a node $v$, an attribute $a$ and an assignment $\rho$,
$E_t^\rho(v,\$x = @a)$ is true if $\rho(x)=\lambda_t^a(v)$. Likewise,
$E_t^\rho(v,\$x \not= @a)$ is true if $\rho(x)\not=\lambda_t^a(v)$.
The semantics of $\semt{p}^\rho$ is then defined accordingly and we
write $t\models^\rho p$ if there is a vertex $v$ of $t$ such that
$(\epsilon,v)\in\semt{p}^\rho$.

\medskip

Deutsch and Tannen \cite{dtxpath} consider a different semantics which does not
assume an external variable binding but rather allows a choice of
values for the variables that makes the expression match. More
formally, $t\models p$ under \defem{existential semantics}, if there is a
variable assignment $\rho$ and a vertex $v$ of $t$ such that
 $t\models^\rho p$. We will write this as $t\models_\exists p$.

We denote the allowance of variables under XPath semantics by \xvars
and under the existential semantics by \evars; the presence of
inequalities with variables is denoted by $\not=$.  For instance,
$\XP(\child,\descendant,\xvars,\ineq)$ denotes the XPath fragment with
\child and \descendant, together with variable equality and inequality
under the XPath semantics.

We define Boolean containment of XP-expressions with attribute values
correspondingly. 
More precisely, under the XPath
semantics, $p\subs q$ holds, if for every tree $t$ and every variable
assignment $\rho$,   $t\models^\rho p$ implies  $t\models^\rho q$. 
 Under the existential semantics, $p\subs q$ if for every tree $t$, if
 there is $\rho$ such that  $t\models^\rho p$ then there is also a
 variable assignment $\pi$ such that    $t\models^\pi q$.

\medskip

In the first subsection, we show that adding variables under XPath
semantics to the basic XPath fragment with disjunction results in a
\xcont problem with {\sc pspace} complexity.

In \cite{dtxpath}, it is shown that \xcont for $\XP(\child,\descendant,
\filter,\wildcard,\disjunction,\evars)$ is 
$\Pi_2^P$-complete~\cite{dtxpath} (Theorems 2.3 and 3.3). Further, they
show that \xcont for
$\XP(\child,\descendant,\filter,\evars)$ is {\sc
  conp}-complete. We show in the second subsection that in the presence of
inequalities these hardness results hold for even smaller fragments.

Adding variables with XPath semantics to the basic fragment with
  disjunction but {\em   without wild-card} gives a $\Pi_2^P$-complete
  \xcont problem. Surprisingly, adding wild-card to this fragment ends
  up in an undecidable \xcont problem.

The results of this section are summarized in Table~\ref{tab:variables}.

For a set $X$ of variables an \defem{equality type} $e$ is an
equivalence relation on $X$. Intuitively, $e$ describes  which
variables have the same value. 

\subsection{XPath semantics}

We start with the {\sc pspace}-upper bound.

\begin{theorem}\label{theo:xvars:upper}
  \xcont for
  $\XP(\child,\descendant,\filter,\disjunction,\wildcard,\xvars,\ineq)$
  is in {\sc pspace}.
\end{theorem}
\proof 
  We basically show that the problem can be reduced to the case
  without variables. Let $p$ and $q$ be two expressions with variables
  $\{x_1,\ldots,x_k\}$ and let $\Sigma'$ be the set of element and
  attribute names which appear in $p$ or $q$ (and which we assume to
  be disjoint). Let $\Sigma=\Sigma'\uplus\{\#\}$. Clearly, $p\subs q$
  holds in general if and only if it holds for trees with element
  names from $\Sigma$.  For each equality type $e$ of the variables
  $x_1,\ldots,x_k$, we construct two expressions $p_e$ and $q_e$ in
  $\XP(\child,\descendant,\filter,\disjunction,\wildcard)$.  By
  construction it then follows that $p\subs q$ if and only if, for
  every equality type $e$, $p_e\subs q_e$. As we can cycle through all
  equality types $e$ and construct each $p_e$ and $q_e$ in {\sc
    pspace}, and each single test will be doable in {\sc pspace}, the
  complexity of the overall algorithm is {\sc pspace}.
  
  Let us first consider the equality type $e$ where all variables are
  pairwise different. With every tree $t$ and every variable
  assignment $\rho$ of type $e$, we associate a tree $t_\rho$ over the
  alphabet $\Gamma= \Sigma\cup\{x_1,\ldots,x_k,\none\}$ as follows. We add, for each attribute $a$ of a node  $v$, a new child of $v$ labeled by $a$ which itself has a child
  which is labeled by one of $x_1,\ldots,x_k$ or with {\tt none}
  depending on $@a$. More precisely, if $@a$ equals some $\rho(x_i)$
  then the grandchild of $v$ is labeled by $x_i$ otherwise by \none.

  Correspondingly, we construct $p_e$ by replacing in $p$ each
  subexpression $\$x_i=@a$ by $./a/x_i$ and each $\$x_i\not=@a$ by
  $./a/(x_1 \mid \cdots \mid x_{i-1} \mid x_{i+1} \mid \cdots \mid x_k
  \mid {\tt none})$. We construct $q_e$ from $q$ in the same way.  It
  is easy to see that, for each $t$, and each variable assignment
  $\rho$ of type $e$, $t\models^\rho p$ if and only if $t_\rho\models
  p_e$.

Thus, it remains to test whether, for all trees of the form $t_\rho$,
$t_\rho\models p_e$ implies $t_\rho\models q_e$.
This can be done along the lines of the proof of Theorem
\ref{theo:finiteupper}. In particular, it is sufficient to consider
only $f(p_e)$-bounded trees. 

This completes the description of the algorithm for equality type
$e$.   If, for some other equality
    type, two variables get the same value then we can 
    replace one of them by the other in $p$ and in $q$. Hence, we get possibly
    fewer variables which are again pairwise different.\qed

We show that the {\sc pspace} upper bound is tight in the general case
and exhibit some lower bounds for restrictions of the formalism.

\begin{theorem}\label{theo:xvars:lower}
  \begin{enumerate}[(a)]
\item \xcont for $\XP(\child,\filter,\xvars,\ineq)$ is \textsc{conp}-hard.
\item \xcont for $\XP(\child,\descendant,\xvars,\ineq)$ is
  \textsc{conp}-hard.
  \item  \xcont for
  $\XP(\child,\descendant,\disjunction,\wildcard,\xvars,\ineq)$
  is {\sc pspace}-hard.

  \end{enumerate}
\end{theorem}
\proof 
  The proofs of (a) and (b) are by reduction from the set of
  unsatisfiable 3SAT-formulas.  To this end let
  $\varphi=\varphi_1\wedge\cdots\wedge\varphi_k$ be a 3CNF formula
  where each $\varphi_i$ is a disjunction $l_{i1}\vee l_{i2}\vee
  l_{i3}$ of 3 literals. Let $x_1,\ldots,x_m$ be the variables
  occurring in $\varphi$.  In both cases, intuitively we construct
  from $\varphi$ an expression $p$ that, describes all (tree
  representations of) possible truth assignments to the literals of
  $\varphi$. The expression $q$ selects a truth assignment if it
  leaves at least one clause unsatisfied. Thus, $\varphi$ is
  unsatisfiable iff $p\subseteq q$.
  
  More precisely, both reductions map $\varphi$ to expressions with
  variables $y$ and $y_1,\ldots,y_m$. The value $\rho(y)$ represents
  $\mytrue$, the other variables represent a truth assignment to
  $x_1,\ldots,x_m$ in the obvious way: $x_i$ is $\mytrue$ iff
  $\rho(y_i)=\rho(y)$.

  In the trees we are interested in, the root $a$-attribute carries
  the data value corresponding to $\mytrue$ and each literal $l_{ij}$
  is represented by one node $v_{ij}$ with an attribute $a$ and label
  $b$. Expression $p$ tests that the attribute values of these nodes
  are consistent with the truth assignment induced by the $y_l$: this
  is checked by variable expressions $v(l_{ij})$ which are $\$y_l=@a$
  if $l_{ij}=x_l$ and $\$y_l\not=@a$ if $l_{ij}=\neg x_l$.

\begin{enumerate}[(a)]
  \item 
In case of  $\XP(\child,\filter,\xvars,\ineq)$, the trees selected by
$p$ have one path of length 3, for each clause. To this end, let, for
each $i$, $p_i$ be the sub-expression
\[
p_i=b[v(l_{i1})]/b[v(l_{i2})]/b[v(l_{i3})],
\]
 and let $p$ be
\[
p=b[\$y=@a][p_1]\cdots[p_k].
\]

Thus, $t\models^\rho p$ guarantees that $t$ has, for each clause
$\varphi_i$, a path in which the $j$-th node ($j=1,2,3$) has an
attribute value consistent with the truth value of $l_{ij}$, as
induced by $\rho$.

Thus, formula $\varphi$ is unsatisfiable, if  every tree which
``passed'' $p$ has a path of length 3 in which all induced truth
values are $\myfalse$. To this end, we define 
\[q=b[\$y=@a]/b[\$y\not=@a]/b[\$y\not=@a]/b[\$y\not=@a].\]

\item Recall that in
  $\XP(\child,\descendant,\vars,\ineq)$ sub-expressions of the forms
  $[\$y=@a]$ and $[\$y\not=@a]$ are allowed. As $\XP(\child,\descendant,\vars,\ineq)$ can not refer to
  branches of a tree, in this case, the vertices $v_{ij}$ have to be
  organized in a linear fashion. 
Basically, in the construction of (a), we replace the
  general $\filter$ subexpressions in $p$ by using $\descendant$ in
  $q$.  To this end, the trees matched by $p$ need to have a path of
  length $4k$ with label pattern $(cbbb)^*$. The $i$-th subpath
  labeled $cbbb$ corresponds to $\varphi_i$.

  Thus, 
  \begin{eqnarray*}
p_i & = & c/b[v(l_{i1})]/b[v(l_{i2})]/b[v(l_{i3})]\\
p & = & b[\$y=@a]/p_1/\cdots/p_k\\
q & = & b[\$y=@a]//c/b[\$y\not=@a]/b[\$y\not=@a]/b[\$y\not=@a]
  \end{eqnarray*}
 By a similar
  reasoning as in (a) it follows that $\varphi$ is satisfiable if
  and only if $p\not\subseteq q$.
\item
 We use basically the same construction as in
    Theorem~\ref{theo:finitelower}. Let $D=\{\sigma_1,\ldots,\sigma_k\}$ be
    the alphabet used in that construction and assume without loss of
 generality that
    $k=2^l$, for some $l$. We use attributes $a_1,\ldots,a_l$ and one
    variable $x$ to encode the symbols of $D$. More precisely,
 $a_1,\ldots,a_l$ and $x$ represent $\sigma_b$, if, the $i$-th digit
 of $b$ is 1 exactly if $x=a_i$. 
The remaining symbols
 $\$$ and $\#$ are still represented by labels. 
 In the expressions $p$ and $q$ the
    element tests are replaced by the wild-card symbol together with
    the respective attribute comparisons.\qed
\end{enumerate}

\subsection{Existential semantics}

Now we turn to \xcont in the context of existential semantics
for variables. In some cases the complexity does not change as to
compared with the XPath semantics, but in others
it raises considerably, even leading to undecidability for  XP(\child,\descendant,
\filter,\wildcard,\disjunction,\evars,\ineq).

\begin{theorem}\label{theo:evars:lower}
\begin{enumerate}[(a)]
\item \xcont for XP$(\child,\filter,\evars)$ is {\sc conp}-hard.
\item \xcont for XP$(\child,\filter,\evars,\ineq)$ is {\sc
    $\Pi^P_2$}-hard.
\item \xcont for $\XP(\child,\disjunction,\evars)$ is
  $\Pi_2^P$-hard. 
\end{enumerate}
\end{theorem}
\proof 
  The proofs for (a) and (b) are similar. Both are reductions from
  containment for Boolean Conjunctive Queries (BCQs). In (b)
  inequalities are allowed, in (a) not. As containment of BCQs and
  BCQs with inequality is hard for {\sc conp}~\cite{cm} and
  $\Pi_2^P$~\cite{vandermeyden}, respectively, (a) and (b) then follow.

We first describe how we represent a relational database $D$ by a
  tree $t_D$. The root is labeled with $S$ and
  for every relation $R$ in $D$ and every tuple $(d_1,\ldots,d_n)$
  in $R$ it has a child labeled $R$ with $n$ attributes
  $@1,\ldots,@n$, where, for each $i$, $@i$ has the value $d_i$.

  A BCQ is a conjunction of relational atoms $R(x_1,\ldots,x_k)$, and,
  in case of (b), inequalities $x\not=y$. Note that atoms of the form
  $x=y$ can be eliminated as each variable must occur in a relational
  atom.  An atom $R(x_1,\ldots,x_k)$ can be represented by the
  subexpression $p_i=R[\$x_1=@1]\cdots[\$x_{k}=@k]$.

We can not represent an
inequality $x\not=y$ by an expression $[\$x\not=\$y]$, as
variables can only be compared with attributes. Nevertheless, as $x$
must occur as the $i$-th entry in a relational atom,
we add $[y\not=@i]$ to the expression for $L$.

We illustrate the construction with an example. For instance, if $Q$
is $E(x,y),E(y,z),x\neq z$ then $p_Q$ is
$$S[E[\$x=@1][\$y=@2][\$z\neq@1] [E[\$y=@1][\$z=@2]].$$

Clearly, for a database $D$, $D\models Q$ if and only if
$t_D\models_\exists p_Q$. On the other hand, from each $t$ (even if it is not
of the intended form) we can define, in a straightforward manner,
using all correctly encoded tuples, a database $D(t)$ such that
$D(t)\models Q$ if and only if $t\models_\exists p_Q$.

Thus, for BCQs $Q_1$ and $Q_2$ it holds $Q_1\subs Q_2$ if and only if
$p_{Q_1}\subs p_{Q_2}$, as required.\\

The proof of (c) is by a reduction from
$\forall\exists$-3SAT~\cite{papad_book}. Note that the fragment
$\XP(\child,\disjunction,\evars)$ allows the use of disjunction of
variable-attribute comparisons.

Let $\varphi=\forall x_1 \cdots x_m \exists
  y_1\cdots y_m \theta$, where
  $\theta=\theta_1\wedge\cdots\wedge\theta_n$ is in 3-CNF with
  variables 
  from $\{x_1,\ldots,x_m,$ $y_1,\ldots,y_m\}$. 
Intuitively, the $\forall\exists$-structure
  of a formula will be mimicked by a {\em for all trees there is a
  variable assignment} statement. Hence, the values for
  $x_1,\ldots,x_m$ will be encoded in the trees, the values for
  $y_1,\ldots,y_m$ will be given by the variable assignment.

  Each tree satisfying $p$ will have two root attributes @\loctrue and
  @\locfalse which represent the truth values \mytrue and \myfalse,
  respectively.  Note that, as there are no inequalities available,
  there will be no guarantee that the values of \loctrue and \locfalse
  are different. We only make use of the node label $b$.  With every
  path of length $m$ which starts immediately below the root we
  associate a (partial) truth assignment as follows: if the
  $c$-attribute of the $i$-th node equals the value of the attribute
  $\loctrue$, we set $\rho(x_i)=\loctrue$, if it is $\locfalse$ then
  $\rho(x_i)=\locfalse$.

The expression $p$ is designed to guarantee the existence of such a path. To this
end, $p$ is $b[\$z_1=@\loctrue][\$z_0=@\locfalse]/p_1/\cdots/p_m$, where, for
each $i$, $p_i=b[\$z_1=@c \text{ or } \$z_0=@c]$. Intuitively, the variables $z_0$
and $z_1$ are used to transport the values \mytrue and \myfalse in the
tree. 

So far, each tree $t$  with $t\models_\exists p$ induces at least one truth
assignment for $x_1,\ldots,x_m$. 

Expression  $q$  checks that  
\begin{itemize}
\item  $z_1$
  and $z_0$ take the values @\loctrue and @\locfalse at the root,
  respectively,
\item   each (XPath) variable $x_i$ has a truth
  value corresponding to the value of the $i$-th node of some path of the
  tree,
\item  each (XPath) variable $y_i$ is either
  @\loctrue or @\locfalse, and 
\item  the induced truth assignment makes $\theta$ true.
\end{itemize}
To this end,
$q=b[e_1]\cdots[e_n][\$z_1=@\loctrue][\$z_0=@\locfalse][\$y_1=@\loctrue \text{
  or } \$y_1=@\locfalse]\cdots [\$y_m=@\loctrue \text{ or }
\$y_m=@\locfalse] /b[\$x_1=@c]/\cdots/b[\$x_m=@c]$, where each $e_i$ is a
variable expression corresponding to $\theta_i$. As an example, if
$\theta_1$ is $x_2\lor y_3 \lor \neg x_1$ then $e_1$ is
$[\$x_2=@\loctrue \text{ or }   \$y_3=@\loctrue  \text{ or }  \$x_1=@\locfalse]$.

Now it is easy to see that $\varphi$ holds iff $p\subseteq q$ with
respect to the existential semantics. Note in particular
that trees for which $@\loctrue=@\locfalse$ at the root fulfill $q$ whenever they
fulfill $p$.\qed

\begin{theorem}\label{theo:evars:upper}
  \xcont for
  $\XP(\child,\descendant,\filter,\disjunction,\evars,\ineq)$ under
  existential semantics is in $\Pi_2^P$.
\end{theorem}
\proof 
We show that for
$\XP(\child,\descendant,\filter,\disjunction,\evars,\ineq)$-expressions
the following holds.
\begin{enumerate}[(a)]
\item $p\not\subseteq q$ if and only if there is a tree $t$ of
size polynomial in $|p|+|q|$ such that $t\models_\exists p$ but $t\not\models_\exists
q$, and
\item $t\models_\exists p$ can be tested in {\sc np}.
\end{enumerate}
Hence, the algorithm {\em Guess a tree $t$ of polynomial size
  and check that $t\models p$ but $t\not\models q$} is a
  $\Sigma_2$-algorithm for the complement of \xcont.

To prove (a), let $p$ and $q$ be expressions and let $p_1 | \cdots | p_m$ and $q_1 |
\cdots | q_n$ be the DNF of $p$ and $q$, respectively.  As the
disjuncts do not contains disjunction themselves they can again be
represented as tree patterns with
additional constraints reflecting the equalities and inequalities
between variables and attributes.

Clearly, $p\not\subseteq q$ if and only if for some
  $i$, $p_i\not\subseteq q$. Hence, as $i$ can be guessed,  in proving
  (a) we can restrict to 
  the case where $p$ does not contain \disjunction. 

We call a tree $t$ \defem{$(p,q)$-canonical} if the following
conditions hold. 
\begin{itemize}
\item The tree structure of $t$ is obtained from the tree pattern $\tau(p)$ 
  by replacing each $\descendant$-edge by a path of length two with
  two child edges and a new intermediate $\#$-labeled node   where
  $\#$ is a label neither occurring 
  in $p$ nor $q$.  Note that the   number of vertices of $t$ is at most
  twice the number of vertices of $\tau(p)$.
\item The attribute values in $t$ are from the set $\{0,\ldots,mk\}$,
  where $m$ is the number of vertices in $t$ and $k$ is the number of
  attributes occurring in $p$ or $q$.
\end{itemize}

Let $S(p,q)$ denote the set of all $(p,q)$-canonical trees.
Note that, as the data values are bounded by $(|p|+|q|)m$ each of these
trees can be encoded by a string of polynomial size.

We show next that, whenever $p\not\subseteq q$ for an expression $p$ from
$\XP(\child,\descendant,\filter,\evars,\ineq)$ and
an expression $q$ from
$\XP(\child,\descendant,\filter,\disjunction,\evars,\ineq)$,
there is a tree $t\in S(p,q)$ that matches $p$ but not $q$.

Therefore let $p\not\subseteq q$ be witnessed by a tree $t'$ not
necessarily from $S(p,q)$. Hence, $t'\models_\exists p$ but $t'\not\models_\exists q$.
Let $e$ be a homomorphism from $\tau(p)$ to $t'$. Let
$a_1,\ldots,a_l$, $l\le k$, be
the pairwise different attribute values of the vertices in $e(\tau(p))$.

We construct  $t$ as follows. Its structure is obtained from
$\tau(p)$ as above by replacing $\descendant$-edges with new nodes
labeled $\#$.  We call a vertex $v$ of $t$ that is already in
$\tau(p)$ an {\em original vertex} and write $p(v)$ for its
corresponding vertex in $\tau(p)$.  An original vertex $v$ of $t$
inherits its attribute values from $e(p(v))$ as follows. If attribute
$b$ of $e(p(v))$ has value $a_i$ then $v$ gets the attribute value
$i$. The attributes of the other nodes get the value 0.

Let $u$ and $u'$ be (not necessarily distinct) original vertices in
$t$ and let $b$, $b'$ be two attributes. Then the 
$b$-attribute of $u$ is different from the $b'$-attribute of $u'$ if
and only if the $b$-attribute of $e(p(u))$ is different from the
$b'$-attribute of $e(p(u'))$.

Clearly, $t\in S(p,q)$ and $t\models p$ via the obvious
homomorphism. It remains to show that $t\not\models q$. Assume
otherwise. Hence, for some $j$, $t\models q_j$. Let $e'$ be a
homomorphism from $\tau(q_j)$ to $t$. As $q$ does not contain the symbol
$\#$ and there are no wild-cards, the image of $\tau(q_j)$ under $e'$ only
contains original vertices of $t$. As these vertices have the same
relationships within each other as their corresponding vertices in
$t'$ we can conclude that $t'$ also matches $q_j$. This concludes the proof of (a).\\

To show (b), we remark that
whether $t\models_\exists p$ for an expression $p$ in
$\XP(\child,\descendant,\filter,\disjunction,\evars,\ineq)$ can
be tested as follows. First, a disjunct  $p_i$ of the disjunctive
normal form of $p$ is guessed. Next, a homomorphism from $\tau(p_i)$ to $t$
and a value assignment for the variables of $p_i$ are guessed (with
values $\le |p_i|$) and it is checked whether all conditions
hold.\qed

We note in passing that for variables with existential semantics even
query evaluation is hard.
\begin{proposition}\label{lem:orvareval}
  \begin{enumerate}[(a)]
  \item 
  Evaluation of Boolean $\XP(\disjunction,\evars)$-expressions is
  \textsc{np}-hard.
\item   Evaluation of Boolean $\XP(\child,\filter,\evars)$-expressions is
  \textsc{np}-hard.
\end{enumerate}
\end{proposition}
\proof 
  \begin{enumerate}[(a)]
  \item 
  The proof is by reduction from 3SAT. Let
  $\varphi=\bigwedge_{i=1}^n\varphi_i$ be a 3CNF formula with variables from
  $\{x_1,\ldots,x_m\}$. Let $t$ be the tree consisting of a single vertex with
  attributes $@ \loctrue=1$ and $@\locfalse=0$. Let $p$ be the expression
  $[\$x_1=@ \loctrue \locor \$x_1=@ \locfalse]\cdots [\$x_m=@ \loctrue \locor
  \$x_m=@ \locfalse] p_\varphi$, where
  $p_\varphi=p_{\varphi_1}\cdots p_{\varphi_n}$ and each
  $p_{\varphi_i}$ represents clause $\varphi_i$. E.g., if
  $\varphi_i=(x_3\lor \neg x_5 \lor x_1)$ then $p_{\varphi_i}$ is
  $[\$x_3=@\loctrue \locor \$x_5=@\locfalse \locor \$x_1=@\loctrue]$. 
Clearly, $\varphi$ is satisfiable iff $t\models_\exists p$.  
\item This follows easily from the correspondence between XP and BCQ as
  explained in the proof of Theorem \ref{theo:evars:lower} and the
  fact that BCQ-evaluation is hard for \textsc{np}.\qed
  \end{enumerate}

The following theorem shows that in the setting of variables with
existential semantics the wild-card has a strong impact. 

\begin{theorem}\label{theo:evars:undec}
\xcont for XP(\child,\descendant,\wildcard,\disjunction,\evars,\ineq)  is undecidable.  
\end{theorem}

\proof 

We use a reduction from Post's Correspondence Problem (PCP) which is
well-known to be undecidable \cite{hu}. An \emph{instance} of PCP is a
sequence of pairs $(x_1,y_1),\ldots, (x_n,y_n),$ where
$x_i,y_i\in\{a,b\}^*$ for $i=1,\ldots,n$. This instance has a
\emph{solution} if there exist $m\in \Bbb{N}$ and
$\alpha_1,\ldots,\alpha_m\in\{1,\ldots,n\}$ such that
$x_{\alpha_1}\cdots x_{\alpha_m} = y_{\alpha_1}\cdots y_{\alpha_m}$.
We construct a DTD $d$, and two XPath expressions $p_1$ and $p_2$ such
that $p_1\subseteq_d p_2$ iff the PCP instance has a solution. At the
end of the proof, we will explain how we can get rid of the DTD.
  
We consider unary XML-trees, that is, strings. They are
roughly of the form $u\$v$, where $\$$
is a delimiter and $u$, $v$ are strings
representing a candidate solution
$(x_{\alpha_1},\ldots,x_{\alpha_m};y_{\beta_1},\ldots,y_{\beta_m})$
for the PCP instance in a suitable way. To check whether such a
candidate is indeed a solution, we roughly have to check whether
\begin{enumerate}
\item for each $i$, $\alpha_i=\beta_i$, that is, corresponding
pairs are taken; and

\item both strings are the same, that is, corresponding
positions in $x_{\alpha_1}\cdots x_{\alpha_m}$ and
$y_{\alpha_1}\cdots y_{\alpha_m}$ carry the same symbol.
\end{enumerate}
To check  (1) and (2), we make use of
a double indexing system based on the values of the attributes block
and position of the nodes in $u$ and $v$. We explain the intuition
behind our reduction by means of a small concrete example.
  
The DTD will define \emph{strings} of the form $S\ u \# v\&$. For
instance, the candidate solution $x_1x_2;y_1y_2$ where $x_1=ab$,
$x_2=b$, $y_1=a$, and $y_2=bb$ will be represented as a concatenation
of $X$- and $Y$-blocks as follows:
$$
\begin{array}{lccccccccc}
&\ S\ &\ X\ &\ 1(x)\ &\ a(x,1,1)\ &\ b(x,1,2)\ &\ X\ &\ 2(x)\ &\ b(x,2,1)\
&\ \#\\%
block&& 1&&&&2\\ position&&&&1&2&&&3
\end{array}
$$ $$
\begin{array}{lcccccccc}
&\ Y\ & \ 1(y)\ &\ a(y,1,1)\ &\ Y\ &\ 2(y)\ &\ b(y,2,1)\ &\ b(y,2,2)\ &\
\&\\ %
block &1&&&2\\ position &&&1&&&2&3
\end{array}
$$
Here, the block and position rows indicate the values of the
respective attributes. The symbols $X$ and $Y$ indicate the beginning
of an $x$- and $y$-block, respectively. The symbol $1(x)$ means that
$x_1$ is picked; and, $a(x,1,1)\ b(x,1,2)$ encode that $x_1$ is the
string $ab$. More precisely, $\sigma(x,i,j)$ encodes that the $j$-th
position in the string $x_i$ is $\sigma$. We need this involved
encoding as we will define a DTD that can only produce valid sequences
of blocks. The attributes ``block'' and ``position'' make up the
double index system as will become clear further on. We refer to
blocks corresponding to encodings of an $x_i$ and $y_i$ as $X$-blocks
and $Y$-blocks, respectively. If a block corresponds to $x_i$ or $y_i$
 we say that its number is $i$.

Let, for each $i\le n$, $x_i$  be $\sigma^{i}_1\cdots \sigma^{i}_{k_i}$
and $y_i$ be $\delta_1^{i}\cdots\delta_{\ell_i}^i$. Then the DTD
$d$  consists of the productions
  $$
\begin{array}{lll} S&\to& X\\ X &\to& 1(x) | \ldots | n(x)\\ \#
&\to& Y\\ Y &\to& 1(y) | \ldots | n(y)\\ \& &\to& \varepsilon
\end{array}$$
and  further sets $P(x_i)$ and $P(y_i)$ of productions,  for each
$i\le n$. Here, $P(x_i)$ 
consists of the productions $i({x})\to \sigma_1^{i}(x,i,1)$, and for
$j:=1,\ldots,k_i-1$, $\sigma_j^i(x,i,j)\to \sigma_{j+1}^i(x,i,j+1)$,
and $\sigma_{k_i}^i(x,i,k_i)\to X\mid\#$.  Analogously,  $P(y_i)$
is the set of productions $i({y})\to \delta_1^i(y,i,1)$, and for
$j:=1,\ldots,\ell_i-1$, $\delta_j^i(y,i,j)\to
\delta_{j+1}^i(y,i,j+1)$, and $\delta_{\ell_i}^i(y,i,\ell_i)\to
Y\mid\&$.  The start symbol is $S$.  Every $X$ and $Y$ has an
attribute \emph{block}; every $a$ and $b$ has an attribute
\emph{position}.  

Formally, a tree $S u\# v\&$ is {\em syntactically correct} if $u$ and
$v$ contain the same number of blocks and it fulfills the following
condition. For $z\in\{u,v\}$, let $\text{block}(z)$ be the list
consisting of the block attribute-values of the nodes in $z$ and let
$\text{position}(z)$ be the list consisting of the position
attribute-values of the nodes in $z$. Then it should be the case that
$\text{block}(u)=\text{block}(v)$ and
$\text{position}(u)=\text{position}(v)$.  A syntactically correct
string $S u\$ v \&$ {\em represents a solution of the PCP instance},
iff the block numbers of corresponding blocks are the same and the
values ($a$ or $b$) of corresponding positions are the same.

Let $p$ be the XP-expression $S$ and let $d$ be as above.
We next construct $q'$ in such a way that it selects the root of an
XML-tree if and only if it is {\em not} syntactically correct or
does {\em not} represent a solution.  As $p$ defines {\em all}
inputs, $p\subseteq_d q'$ if and only if the PCP instance has
\emph{no} solution.

In the following, if
$z$ is the string $abab$ then we use $z$ as a shorthand for $a/b/a/b$.
Further, we denote the string generated by the grammar from $x_i$ and
$y_i$ by $\tilde x_i$ and $\tilde y_i$, respectively.
That is, for $x_1=ab$, $\tilde{x_1}=a(x,1,1)b(x,1,2)$.

\begin{enumerate}

\item The block index is wrong.
\begin{enumerate}
\item the block value of the first $X$ in $u$ differs from the block value
of the first $Y$
in $v$: $$ S/X[\$d = @block]/\!/\#/Y[\$d \neq @block].$$

\item the block value of the last $X$ in $u$ differs from the block value of
the last $Y$
in $v$: for each $i,j\in\{1,\ldots,n\}$ we have 
 $$ S/\!/X[\$d=@block]/i(x)/\tilde x_i/\#/\!/Y[\$d\neq @block]/j(y)/\tilde y_i\&$$

\item two $X$-block values are the same:
$$ S/\!/X[\$d=@block]/\!/X[\$d=@block]//\#$$

\item two $Y$-block values are the same;
$$  S/\!/\#/\!/Y[\$d=@block]/\!/Y[\$d=@block]$$

\item two successive $X$-block values are not successive in $v$:
  for all $i,j\in\{1,\ldots,n\}$ we have \\
  $S/\!/X[\$d=@block]/i(x)/\tilde x_i/
  X[\$e=@block]/\!/$\hfill{}\\
  \phantom{1}\hfill $\# {}/\!/Y[\$d=@block]/j(y)/\tilde y_j/Y[\$e\neq
  @block].$ 

\end{enumerate}

\item The position index is wrong. This is done in an analogous fashion.

\begin{enumerate}
\item the first position in $u$ differs from the first position in
$v$:

$$ S /X/*/*[\$d=@position]/\!/Y/*/*[\$d\neq @position].$$

\item the last position in $u$ differs from the last in $v$:
$$ S/\!/ *[\$d=@position]/\#/\!/*[\$d\neq @position]/\&$$

\item two $X$-position values are the same:
$$ S/\!/*[\$d=@position]/\!/*[\$d=@position]//\#$$

\item two $Y$-position values are the same;
$$  S/\!/\#/\!/*[\$d=@position]/\!/*[\$d=@position]$$

\item two successive $X$-position values are not successive in $v$:
we have to deal with several cases as the successive positions 
can occur in the same block or in successive blocks.
\begin{enumerate}
\item the $X$-positions occur in the same block, the $Y$-positions
occur in the same block:
for all $i,j\in\{1,\ldots,n\}$, $k\in\{1,\ldots,|x_i|-1\}$,
$\ell\in\{1,\ldots,|y_j|-1\}$:

\begin{multline*} S/\!/X/i(x)/*^{k-1}/*[\$d=@position]/*[\$e=@position]
\\/\!/Y/j(y)//*^{\ell-1}/*[\$d=@position]/*[\$e\neq @position]
\end{multline*}

\item the $X$-positions occur in successive blocks, the $Y$-positions
occur in the same block:
for all $i,j\in\{1,\ldots,n\}$,
$\ell\in\{1,\ldots,|y_j|-1\}$:
\begin{multline*} S/\!/X/i(x)/*^{|x_i|-1}/*[\$d=@position]/
X/*/*[\$e=@position]
\\/\!/Y/j(y)//*^{\ell-1}/*[\$d=@position]/*[\$e\neq @position]
\end{multline*}

\item the $X$-positions occur in the same block, the $Y$-positions
occur in successive blocks:
for all $i,j\in\{1,\ldots,n\}$,
$k\in\{1,\ldots,|x_i|-1\}$:

\begin{multline*} S/\!/X/i(x)/*^{k-1}/*[\$d=@position]/
*[\$e=@position]
\\/\!/Y/j(y)//*^{|y_j|-1}/*[\$d=@position]/Y/*/*[\$e\neq @position]
\end{multline*}

\item the $X$-positions occur in successive blocks, the $Y$-positions
occur in successive blocks:
for all $i,j\in\{1,\ldots,n\}$:
\begin{multline*} S/\!/X/i(x)/*^{|x_i|-1}/*[\$d=@position]/
X/*/*[\$e=@position]
\\/\!/Y/j(y)//*^{|y_j|-1}/*[\$d=@position]/Y/*/*[\$e\neq @position]
\end{multline*}
\end{enumerate}
\end{enumerate}

\item  $w$ does not represent a solution:
\begin{enumerate}
\item The block number for some block in $u$ is different from
the corresponding block in $v$: for all $i,j\in\{1,\ldots,n\}$
with $i\neq j$:$$
S//X[\$d=@block]/i(x)/\!/Y[\$d=@block]/j(y)$$

\item The symbol ($a$ or $b$) in $u$ is different from
  the corresponding symbol in $v$. Thereto, we have the
following expressions: for all 
$i,j\in\{1,\ldots,n\}$, $k\in\{1,\ldots,|x_i|\}$,
$\ell\in\{1,\ldots,|y_j|\}$ 
$$
S//a(x,i,k)[\$d=@position]/\!/\#/\!/b(y,j,\ell)[\$d=@position]$$
and
$$
S//b(x,i,k)[\$d=@position]/\!/\#/\!/a(y,j,\ell)[\$d=@position]$$
\end{enumerate}
\end{enumerate}
Clearly, $w$ is not syntactically correct or does not represent a
solution if and only if one of the above conditions holds.

To get rid of the DTD, we add disjuncts to $q'$ that capture 
violations of the DTD. However, to this end we need to express that
children of a node cannot have a certain label. As we can not express
this kind of negation directly, we encode labels of nodes by equality
types of attribute values. So, let $L:=\{\ell_1,\ldots, \ell_m\}$ be
the set of all the labels we need.  Every node now has $m$ attributes
$a_1,\ldots,a_{m}$. If for a node, $j\ge 1$ is the largest number such
that the value of $a_1$ equals the value of $a_j$ then the node is
considered as labeled with $\ell_j$. One can
match a node labeled with $\ell_j$ by checking the corresponding
equality type: for instance, by an expression of the form
$$
*[\$x_1=@a_1][\$x_2\neq @a_1]\cdots[\$x_{j-1}\neq @a_1]
[\$x_j= @a_1].
$$
Clearly, when using this approach we can express that a certain
node is not labeled by a certain label. For every rule $a\to b_1\mid
\cdots\mid b_k$ in the DTD we add the disjunct $//a/c$ to $q'$ where
$c\in L\setminus \{b_1,\ldots,b_k\}$.  When we write $//a/c$, we of
course mean XPath expressions taking labels into account as specified
in the manner above.  Let $q$ be obtained from $q'$ by adding all
the disjuncts from the DTD and replacing all references to labeling by
references to encoding with attributes. Note that now we allow
arbitrary trees as well.

It remains to argue that $p\not\subseteq q$ iff the PCP instance
has a solution.  When there is a solution to the PCP then clearly the
encoding of this string will match $p$ but not $q$. Suppose that
there is a tree that matches $p$ but not $q$. This means that no
error occurs on any path in the tree. Therefore, every path is an
encoding of a solution to the PCP instance.\qed


\section{Discussion}
\label{sec:discussion}

The article studied the complexity of the containment problem for a
large class of XPath expressions. In particular, we considered
disjunction, DTDs and variables. Unfortunately, the complexity of
almost all decidable fragments lies between {\sc conp} and {\sc
  exptime}. On the other hand, the size of XPath expressions is rather
small. As pointed out, Deutsch and Tannen, and Moerkotte already
obtained undecidability results for XPath containment. We added modest
negation ($\neq$) and variables with the existential semantics. In
\cite{NevenSXPath03} a corresponding result is shown in the presence
of node-set equality.  However, the reduction employs XPath expressions
with absolute filter expressions which fall outside the scope of the
present paper.  It would be interesting to have a precise
classification of which combination of features makes the problem
undecidable.

Although the complexity has been settled for a lot of fragments of
XPath with child and descendant axes, the picture is by no means
complete. A particular case that remains open is the case of
$\XP(\dtd,\child,\descendant,\wildcard)$.

Corresponding results in the presence of other axes remain to be
investigated. Besides the general upper bound of Marx \cite{marxedbt}
for navigational XPath with all axes, few precise results are
known. Work on satisfiability of XPath with the sibling axis has been
done by Fan and Geerts~\cite{DBLP:conf/dbpl/GeertsF05}.

\section*{Acknowledgment}
We thank Stijn Vansummeren for comments on a previous version of this
paper. We thank the anonymous referees of our ICDT 2003 paper and the
referees of this article for many valuable suggestions.


\bibliographystyle{plain}
\bibliography{database}

\end{document}